\documentclass[letter]{jpsj3}
\usepackage{txfonts}
\usepackage{color}

\title{
Important Descriptors and Descriptor Groups of Curie Temperatures of Rare-earth Transition-metal Binary Alloys
}

\author{
Hieu Chi Dam$^{1,2,3}$, Viet Cuong Nguyen$^4$, Tien Lam Pham$^{1,5}$, Anh Tuan Nguyen$^6$,  \\
Kiyoyuki Terakura$^{1,2}$, and Takashi Miyake$^{2,5,7}$, and Hiori Kino$^{2,5}$ 
}
\inst{
$^{1}$Japan Advanced Institute of Science and Technology, 1-1 Asahidai, Nomi,Ishikawa 923-1292, Japan\\
$^{2}$ CMI$^2$, MaDIS, NIMS, Tsukuba, Ibaraki 305-0047, Japan\\
$^{3}$JST, PRESTO, 4-1-8 Honcho, Kawaguchi, Saitama, 332-0012, Japan\\
$^{4}$HPC Systems Inc., Japan\\
$^{5}$ESICMM, NIMS, Tsukuba 305-0047, Japan \\
$^{6}$Hanoi Metropolitan University, 98 Duong Quang Ham, Cau Giay, Hanoi, Vietnam \\
$^{7}$CD-FMat, AIST, Tsukuba, Ibaraki 305-8568, Japan
}

\abst{
We analyze the Curie temperatures of rare-earth transition metal binary alloys using machine learning. 
{
In order to select important descriptors and descriptor groups, we }introduce a newly developed subgroup relevance analysis and adopt hierarchical clustering in the representation. 
We execute exhaustive search and demonstrate that our approach results in the successful selection of important descriptors and descriptor groups. 
It helps us to choose the combination of descriptors and to understand the meaning of the selected combination of descriptors.
}

\def\TC{$T_C$}
\def\TCs{$T_C$s}
\def\ElecnegT{$\chi_T$}
\def\covrT{$r_{T}^{cv}$}
\def\covrR{$r_{R}^{cv}$}
\def\dRT{$d_{TR}$}
\def\dTT{$d_{TT}$}
\def\IPT{$IP_T$}

\begin{document}
\maketitle

Magnets are now widely used and play an important role in energy savings.{\cite{sugimoto11,hirosawa17}} One of the most important applications of magnets is electric motors, whose performance significantly depends on the performance of magnets. Nd-Fe-B based rare-earth magnets are the strongest among the existing permanent magnets, and are almost the only type of permanent magnets that meets the stringent performance requirements of the recent electric motors. However one of the problems with Nd-Fe-B magnets is the relatively low Curie temperature compared to the operation temperatures of the motors. Therefore, many researchers have carried out studies to overcome this drawback, including the exploration of new magnets. 

  The Curie temperature (\TC) is one of the most important physical quantities of magnets, {but} unfortunately, it is one of the most difficult physical quantities to predict correctly. There are several theory-driven methods for evaluating the \TC\ of magnetic materials. \cite{miyake18} One of the basic approaches is to solve an (extended) Hubbard model by using various low-energy solvers. In principle, this method is expected to be accurate. However Anisimov et al. showed that the results are sensitive to the effective parameters and details of the low energy solver. \cite{Anisimov2013,Anisimov2014,Anisimov2016} Therefore, this approach is still at the level of testing the formalism for simple systems like pure transition-metal magnets.
  
  Atomistic spin model is the most common choice for practical application to more complex systems. \cite{miyake18} The spin model is constructed from the magnetic moment at each atomic site and the intersite magnetic exchange-couplings based on the assumption of fixed magnitude of spin moments. The parameters are evaluated using the first-principles calculations. \cite{miyake18} This method can be applied to rare-earth magnets. Usually, the model is simplified further, and is restricted to the TM-3d and RE-4f spins. Then, \TCs is evaluated, usually in the mean field approximation. The mean field approximation, however, usually overestimates \TCs. Thus, there exist many sources of error in the {\TC} evaluation using the atomistic spin model. The development of theoretical methods for the estimation of the \TC\ is still underway.
  
   In contrast to the deductive approaches described so far, there is now a movement toward utilizing inductive approaches, i.e., data-driven methods for estimating \TC, and there have been {many reports of} successful prediction of the physical {quantities} using such methods.{\cite{ Potyrailo2011,Rajan2013book,Agrawal2016,Jain2016,Liu2017,Lu2017} } The data-driven approach accumulates data, prepares descriptors, makes {a model with the descriptors}, and finally predicts the values of physical quantities of new materials. One of the key points to be considered for successful prediction is the choice of descriptors. A typical example of descriptor selection can be seen in the work by Ghiringhelli et al., where a {regression} model is used to predict the energy difference between zinc blende or wurtzite and rocksalt structures. \cite{Ghiringhelli15} They used a linear regression model, and first prepared basic descriptors. However, a linear regression model with only the basic descriptors has low description power. Then, they performed various operations on the basic descriptors and produced a number of nonlinear combinations among the basic descriptors.  This resulted in an increase in  the prediction power. They shrank the number of descriptors using LASSO and finally employed exhaustive search to find the best linear {regression} model.  Their work shows that the combination of descriptors is important for increasing the accuracy of the { regression} model.
  
   {Usually, } we select the best {regression} model and discard all the others (performance-optimized model). {However} we know that there exist many {regression} models, where the combination of the descriptors is different from the one that has the best score, but the score of which is as good as the best one indicated by the exhaustive search method. { (The best score means, for example, the largest $R^2$ value in the regression model.)} There exists another strategy where we choose the { regression} model the score of which is not the best, but is high. {For example, we can choose low cost descriptors, where "low cost"} means easy or literally low cost to evaluate through experiments or calculations. {This model} is usually referred to an operation-optimized model. Okada et al. devoted considerable effort to the latter problem. They showed the scores of {regression} models as the density of states to understand the overall structure in one way, and plotted the best scores as a function of the combinations in another way, such as the indicator diagram, to select the best combinations depending on the purpose of the analysis.\cite{Okada1,Okada2,Okada3} 

Yet, it is not easy to understand the relationship and structures among descriptors from a huge list of scores and descriptors. 
Informatics treatment usually ignore the importance of the meaning of the descriptors, though they are physical parameters that physicists { regard} as important. 
{However} we hope that we can extract more information from the huge data. 
In the present work, we introduce a well-defined subgroup concept to clarify the relationship among descriptors. Our method can also elucidate how to choose combination of descriptors systematically as well as how to understand the meaning of descriptors.

\begin{table}

\begin{tabular}{p{2.5cm}|p{5.0cm}}
\hline\hline
\centering
Category &Descriptors \\
\hline
Atomic properties of transition metals  (T) & $Z_T$, $r_T$, $r^{cv}_{T}$, $IP_{T}$, $\chi_{T}$, $S_{3d}$, $L_{3d}$, $J_{3d}$\\
\hline
Atomic properties of rare-earth metals  (R)  & $Z_R$, $r_{R}$, $r^{cv}_{R}$, $IP_{R}$, $\chi_{R}$, $S_{4f}$, $L_{4f}$, $J_{4f}$, $g_J$, $J_{4f}g_{J}$, $J_{4f}(1-g_J)$\\
\hline
Structural information (S)  &$C_T$, $C_R$, $d_{T-T}$, $d_{T-R}$, $d_{R-R}$, $N_{T-R}$, $N_{R-R}$, $N_{R-T}$\\
\hline\hline
\end{tabular}
\caption{\label{descriptorDefinition} Transition metal, rare-earth, and structural descriptors. See also the supporting information.\cite{supportinginfo}}
\label{descriptorDefinition}
\end{table}

  Our target variable is the experimental \TC\ of the rare-earth transition-metal binary stoichiometry alloys considered in this study. \cite{atomwork} We select the descriptors from the element dependent categories (R for rare-earth elements and T for transition metal elements), and utilize the knowledge of the conventional theory-driven method. The key parameters of the effective theory-driven models are related to the properties of the constituent elements and/or structural parameters. For example, the orbital energy level increases (becomes deeper) as the atomic number $Z$ increases. The electron interaction becomes stronger as the atomic orbital becomes more localized. The magnetic exchange-couplings are associated with the strength of the electron interaction and transfer integrals. The coupling strength between TM-3d and RE-4f (through RE-5d) is crucial for discussing the RE dependence of magnetism. This strength is proportional to the 3d-4f effective exchange coupling and the 4f total spin projected onto the 4f total angular moment $J_{4f}$. The latter quantity is given by $J_{4f}(1-g_J)$, with $g_J$ being the Land\'e g-factor. We also add the descriptors from the structure-related category (S) to describe the ratio of the elements as well as the real volume or spatial dependent simple variables to distinguish, e.g., Th$_2$Zn$_{17}$ and Th$_2$Ni$_{17}$ polytypes. We list the descriptors in Table~\ref{descriptorDefinition}, and give their detailed explanations in the supporting information.\cite{supportinginfo}
  
 As a {regression model}, we employ kernel ridge regression with the radial basis function kernel. Kernel ridge regression can include the non-linear effects of the descriptors and has much stronger power to fit the target functions with the descriptors, though there exist a demerit of taking much more time to fit/predict the {regression} models than the linear regression does. We used Python scripts with mpi4py, scipy and scikit-learn.\cite{scipy1,scipy2,sklearn} Our scores in the {regression} models are the $R^2$ values, which we evaluate in the leave-one-out cross validation.

First, we analyze the descriptors. We take Pearson's correlation coefficient between the descriptors. For the T category, the absolute values of Pearson's correlation coefficient among the three descriptors, $Z_T$, $r_T$, and $S_{3d}$, are the same, namely 1, which means that their contributions are the same in the {{regression} model after the normalization procedure. Therefore, the number of independent descriptors is reduced from 27 to 25. Then, we perform exhaustive search for $2^{25}-1=3.3\times 10^{7}$ {regression} models where the combinations of descriptors are different, and evaluate their accuracy values (scores). 
  
  Usually, we evaluate the score of the {regression} model; however, we want to evaluate the importance of the descriptors. Therefore, we change the viewpoint from the {regression} model to the descriptor in order to discuss the importance of the latter. We use relevance analysis, \cite{RelevanceAnalysis1, RelevanceAnalysis2} which roughly corresponds to the linear response theory with respect to the descriptors. ({ We explain the scores and relevance analysis in} the supporting information.\cite{supportinginfo}}) It originally utilizes the change in values when we remove/add {\it a descriptor}. The former corresponds to the leave-one-out experiment, while the latter corresponds to the add-one-in experiment. The descriptor is strongly or weakly relevant when its accuracy score changes meaningfully in the leave-one-out or the add-one-in experiment, respectively.  

\begin{figure}
\begin{center}
\includegraphics[width=8cm]{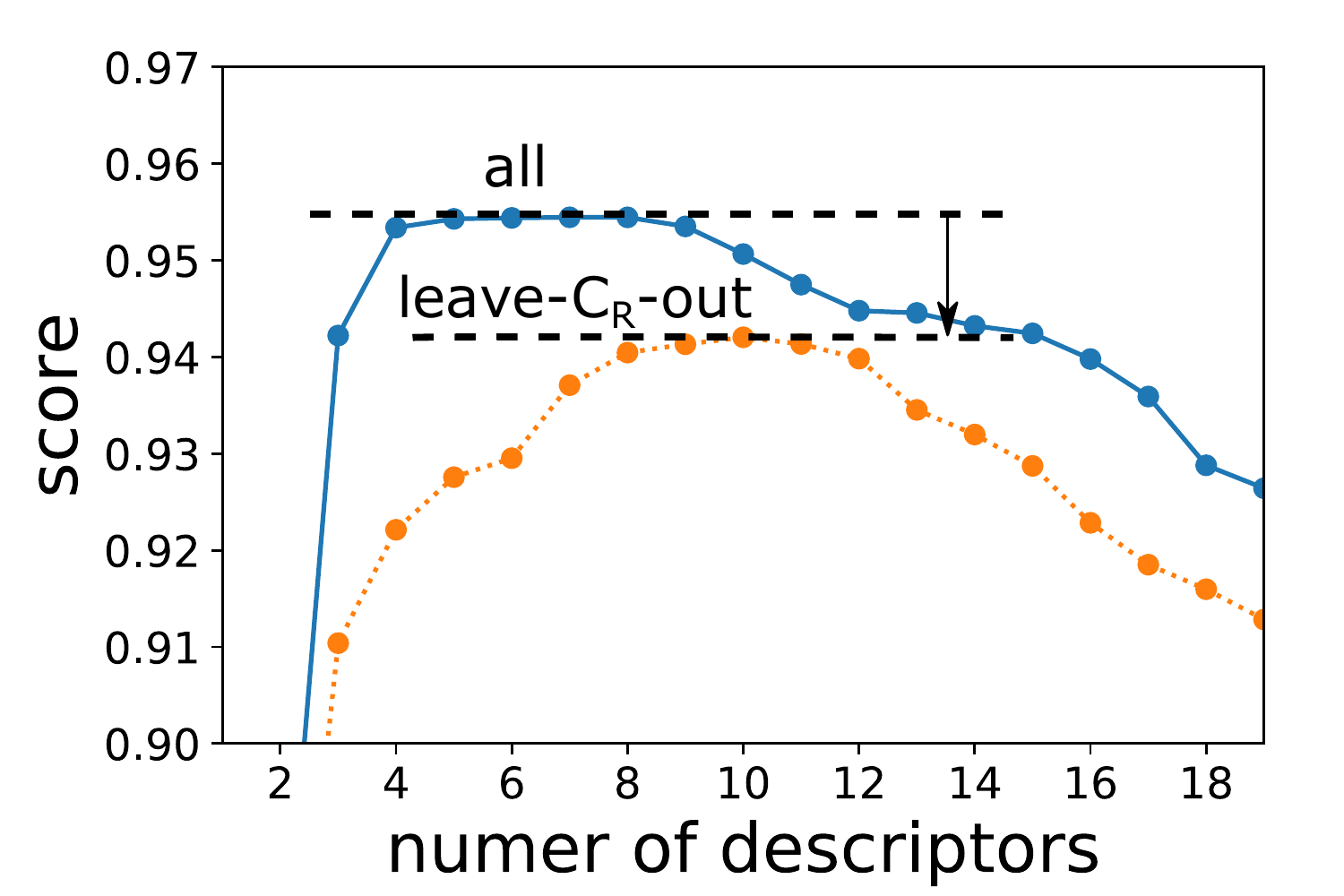}\\
\includegraphics[width=8cm]{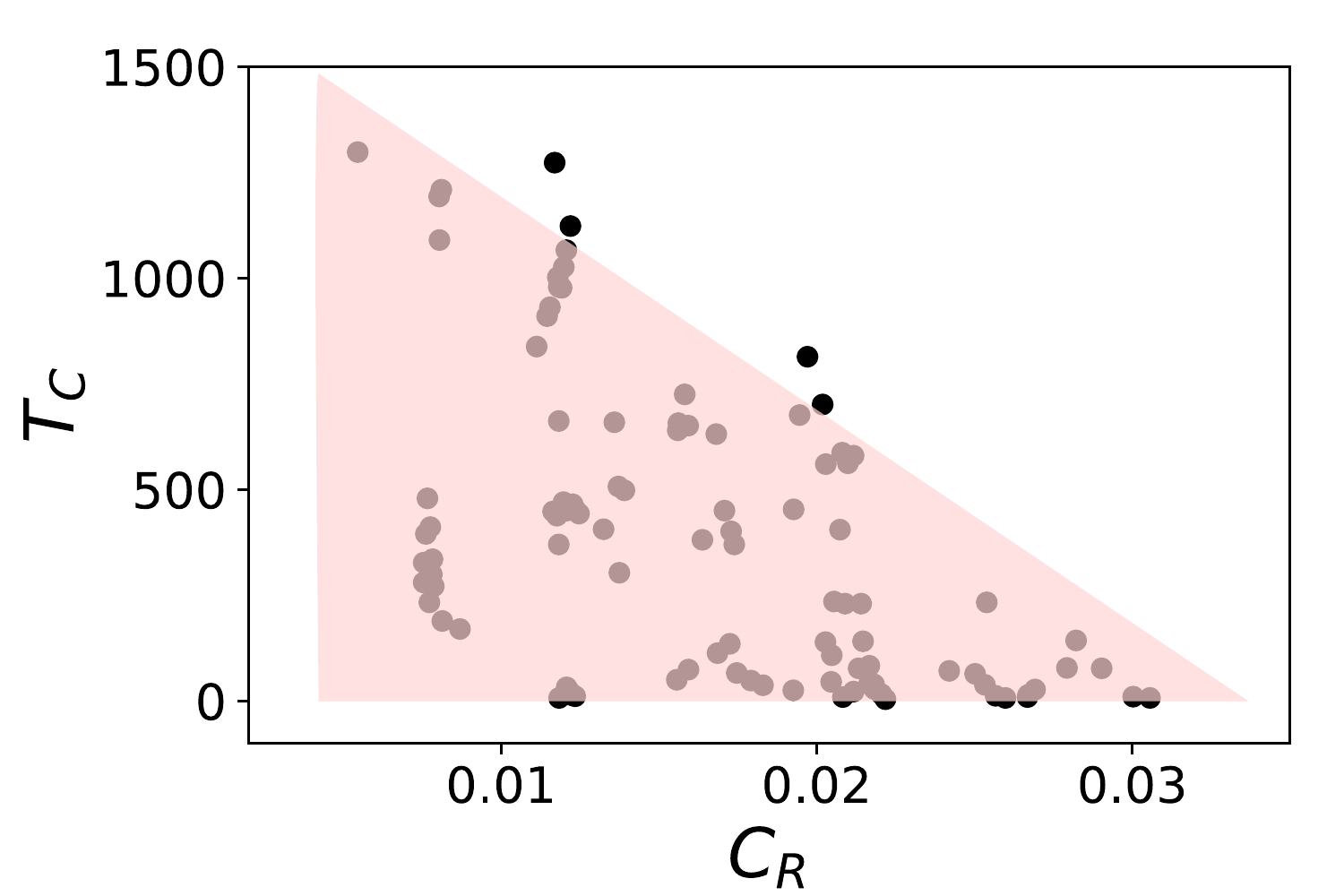}
\end{center}
\caption{(Color online) \label{numberAccuracy} Top panel: The blue line shows the best score for each number of descriptors. The orange dotted line shows the score when $C_R$ is removed. Bottom panel: $C_R$ ($\AA^{-3}$) vs \TC\ ($^\circ$C).  }
\label{numberAccuracy}
\end{figure}
  
  Our first relevance analysis is based on strong relevance. We found that only the descriptor, $C_R$, is strongly relevant. We can verify the importance of $C_R$ when we plot $C_R$ vs \TC. Almost all the points are placed in the bottom-left side of the right panel of Fig.~\ref{numberAccuracy}. Thus, it is clear that $C_R$ has a considerable influence on the \TC. It should be noted that that we will not able to find such a relationship if we simply execute the {regressions}. 
 
{We notice that relevance analysis can be done not only for a descriptor, but also for a subgroup of descriptors. We define groups and subgroups in this paragraph. }
  The second relevance analysis is based on weak relevance, where, in the original prescription, we add another {\it descriptor} to the set of descriptors, which we must define. We define the groups and subgroups here, and make use of them in the relevance analysis. We utilize hierarchal clustering analysis, where the distance between descriptors is one minus the absolute values of Pearson's correlation coefficient. {We can define the groups or subgroups of descriptors that are clustered based on the criteria of them being within distance, $d$, of each other.} For example, we can define four groups at $d = 0.5$. Two of them have the same descriptors as those of the T and R categories, while the other two have that of the original S category. (We call the original cluster as {\it category} and the cluster by the hierarchical analysis as {\it group}.) The \dRT\ constitutes a group, while the other S category descriptors constitute the other. It is not surprising that the grouping at $d =0.5$ is almost the same as the categories defined a priori as T, R, and S when we remember the definition of the descriptors of the materials. Here, we successfully defined the groups and subgroups, where the groups are almost the same as the original category but are clustered from the data themselves. (We redefine the group S as a result of this clustering. The group S that does not include \dRT\ is different from the category S.)

\begin{figure}
\includegraphics[width=17cm]{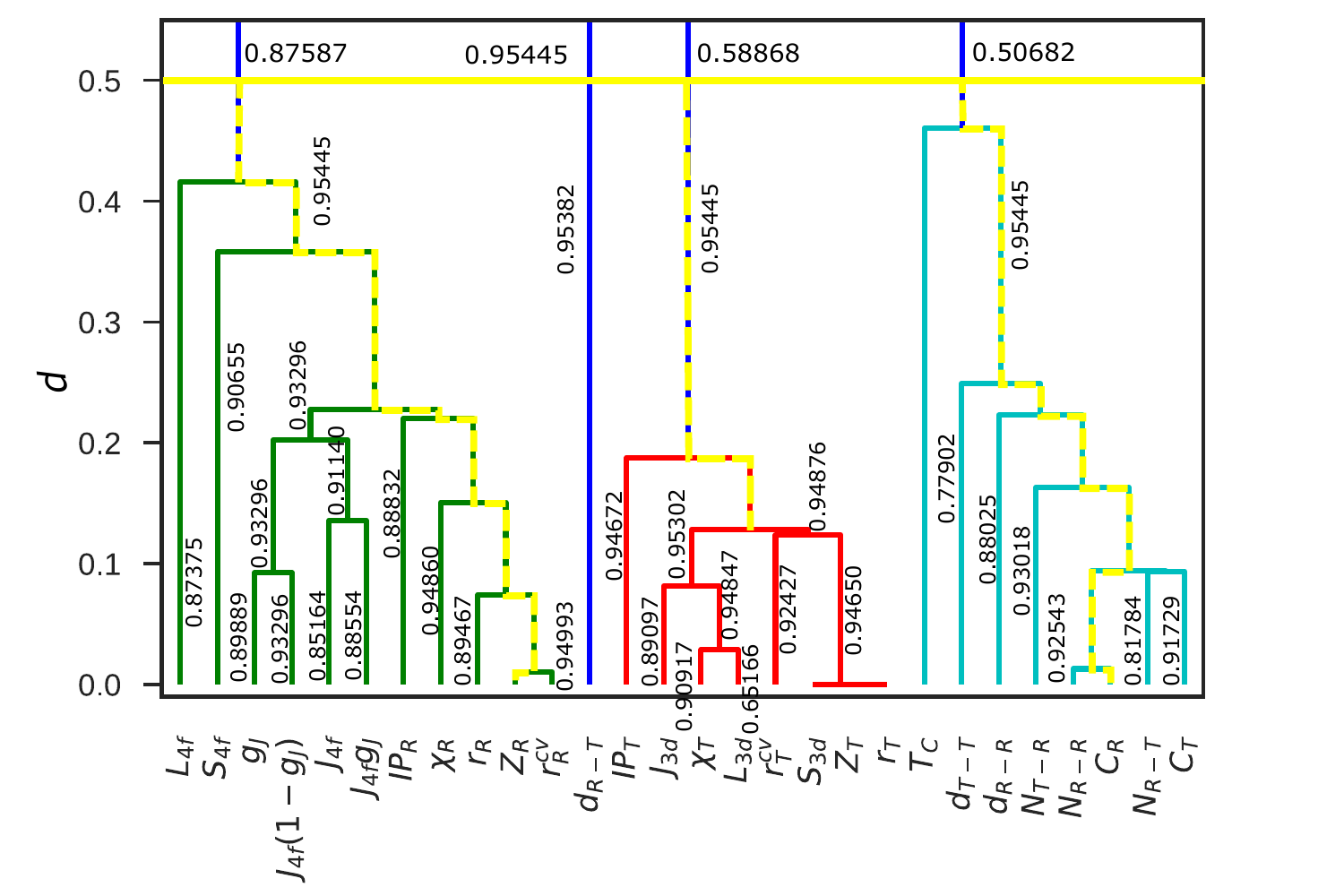}
\caption{(Color online) \label{subgroupRelevanceAnalysis} $R^2$ scores of the subgroup relevance analysis on the hierarchical clustering of the descriptors. 
We include \TC\ in the dendrogram.
The group R (green) is from $L_{4f}$ to \covrR. The group T (red) is from \IPT\ to $r_T$. The group S (cyan) is from \dTT\ to $C_T$. 
 The group \dRT\ is made of the descriptor \dRT. 
The horizontal values are strong relevance values and the tilted values are weak relevance values. The vertical axis {shows} the distance, $d$, and the values are one minus the absolute values of Pearson's correlation coefficient. The paths of the highest value (0.9544{5}) are colored in yellow dashed lines. See details in the main body also.} 
\label{SubgroupRelenvanceAnalysis}
\end{figure}

  We can make further advances in this grouping. We notice that the definition of the value of $d$ is unnecessary, but we only have to define the vertical line of the decomposition tree to define the subgroups because the child nodes below the vertical line is the same. {(See also Fig.~\ref{SubgroupRelenvanceAnalysis}. The vertical axis corresponds to $d$.)} Thus, we are able to define many subgroups of the descriptors as sets of the child nodes of the {\it dendrogram}. 
  
  {We} apply the relevance analysis not  to a descriptor but to a subgroup/group. We call this method {\it subgroup relevance analysis}. We plotted the result in Fig.~\ref{SubgroupRelenvanceAnalysis}. The horizontal score is evaluated in the leave-one-out experiment and is related to the strong relevance, while the vertical scores are evaluated in the add-one-in experiment and  is related to the weak relevance. Note that the score of a subgroup belonging to {a} group is evaluated under the condition that we must use at least one descriptor in the subgroup, and any descriptors belonging to the other group{{s}} can be added in the weak relevance analysis.
  
  In Fig.~\ref{SubgroupRelenvanceAnalysis}, the weak relevance values, or {add-one-in} values, are written as vertical values. The subgroup containing only $r_R$ has the score, 0.8946{7}, which is the highest score in the condition that we must take the subgroup $r_R$ in the group R and we can take any descriptors in the other groups. (A subgroup which has a descriptor is also a subgroup.) The subgroup containing $r_R$, $Z_R$, and \covrR\ has the score, 0.9544{5}, which is the highest score in the condition that we must take at least one descriptor in the subgroup $r_R$, $Z_R$, and \covrR\ of the group R and we can take any descriptors in the other groups as explained in the previous paragraph. 

\begin{table}
\caption{ \label{nvsR2table} The best $R^2$ score and descriptors as a function of the number of descriptors $n$. }
\label{nvsR2table}
\begin{tabular}{|c|c|l|} \hline
n&score&descriptor(s) \\ \hline

2&0.8701{ 5}& $C_R$,$Z_T$ \\
3&0.9422{ 2}& $C_R$,$Z_R$,$Z_T$ \\
4&0.9533{ 9}& $J_{3d}$,$C_R$,$Z_R$,$Z_T$ \\
5&0.9542{ 9}& $L_{3d}$,$J_{3d}$,$C_R$,$Z_R$,$Z_T$ \\
6&0.9543{ 9}& $L_{3d}$,$J_{3d}$,$\chi_T$,$C_R$,$Z_R$,$Z_T$ \\
7&0.9544{ 5}& $L_{3d}$,$J_{3d}$,$\chi_T$,$C_R$,$Z_R$,$Z_T$,\covrT \\
8&0.9544{ 5}& $L_{3d}$,$J_{3d}$,$\chi_T$,\IPT,$C_R$,$Z_R$,$Z_T$,\covrT \\
\hline
\end{tabular}

\end{table}
  
  The sole descriptor $Z_R$ in the group R has the highest score (0.9544{5}). It means that $Z_R$ can solely represent the group R. This is also the case for the $C_R$ subgroup in the group S. {However} the structure of the group T is different from those of the groups R and S. The subgroup made of $J_{3d}$, \ElecnegT, \covrT, $Z_T$ (and $r_T$ and $S_{3d}$) has the highest score (0.9487{6}), but its child subgroup descriptors { have} smaller scores (0.9242{7} and 0.9465{0}). It means that there exists no single descriptor that can represent the overall nature of the group T. When we examine all the combinations made of $J_{3d}$, \ElecnegT, \covrT, $Z_T$, we find that $Z_T$ takes the best score (0.9545{0}) if we choose { only one of the } descriptors among them, a set of $Z_T$ and $J_{3d}$ is the best (0.9533{9}) for two descriptors, and a set of $Z_T$, $J_{3d}$ and $L_{3d}$ is the best (0.9544{5}) for three descriptors. { We note that the descriptor $Z_T$ has the same effect as $S_{3d}$. We discuss interpretation of the result later. }
  
  We can also obtain the importance of the groups from the horizontal values above the yellow solid line in Fig.~\ref{SubgroupRelenvanceAnalysis}. They are the strong relevance values, or leave-one-out values of the groups T, R, and S. For example, the group R has the value, 0.8758{7}, which is the best score when we remove all the descriptors of the group R. The better the score is, the less important the group is. The value, 0.5068{2}, is the smallest among them, which means that the group S is the most important among the groups. On the other hand, the least important group is R, the value of which is 0.8758{7}. It means that the score still holds a high value even if we exclude all the descriptors in the group R. Therefore, the importance of group R is the lowest among T, S, and R.

  We have added additional explanation in Fig.~\ref{SubgroupRelenvanceAnalysis}. The descriptor $J_{4f} (1-g_J)$ can represent the subgroup containing $g_J$,...,$J_{4f} g_J$, but the score is 0.9329{6}, which is lower than the score 0.9544{5} of $Z_R$. We have also added a comment on the group of \dRT. The strong relevance value is 0.9544{5} and the weak relevance value is 0.95382. The facts that their difference is small and that the weak relevance value is smaller than the strong relevance value mean that the existence of the group \dRT\ makes the { regression} model worse. 

  Here, we compare the result of the subgroup relevance analysis shown in Fig. \ref{SubgroupRelenvanceAnalysis} with the best score having $n$ descriptors without the subgroup relevance analysis, which is shown in table \ref{nvsR2table}. The set of $C_R$, $Z_R$, and $Z_T$ has the best score ({0.94222}) for $n=3$. The set of $C_R$, $Z_R$, $Z_T$, and $J_R$ has the best score (0.9533{9}) for $n=4$. The set of $C_R$, $Z_R$, $Z_T$, $J_R$, and $L_{3d}$ has  the best score (0.9542{9}) for $n=5$. The descriptor sets are made of the most important descriptors in group R ($Z_R$), group S ($C_R$), and group T ($Z_T$ when we choose a descriptor; $J_{3d}$ and $Z_T$ when we choose two descriptors; and $J_{3d}$, $L_{3d}$, and $Z_T$ when we choose three descriptors.) These combinations are the same as the analysis in the previous paragraph. Thus, the subgroup relevance analysis successfully illustrates the structure among the descriptors and their importance. 
  
{One may think that the difference in the scores are quite tiny. For example, 99.0\% value of the global best score is 0.944, which roughly corresponds to the best score with 12 descriptors (see also Table I in the supporting information). \cite{supportinginfo} However the predicting ability changes drastically. We plot the "RMSE" between the best models with $n$ descriptors in Fig. 2 in the supporting information.\cite{supportinginfo} It can be clearly seen that the prediction abilities for  $n$=3 to 8 is qualitatively different from those for $n\ge$9, but the difference of the score of the best model with 9 (10) descriptors to the global best model is only 0.1\% (0.4\%). The difference in the score looks tiny at a glance, but is meaningful in this data and regression model. (One must also discuss the total density of state of the scores to discuss the meaningful difference of the scores, but it is beyond the scope of this study.\cite{Okada1,Okada2,Okada3})

The ordering of the scores of the models (combinations of descriptors) can be changed according to the details of the regression scheme and noise in the data, because the differences in the scores are quite small (Table II in the main body and Table I in the supporting information).\cite{supportinginfo} Thus, just showing the best models with $n$ descriptors may give us wrong information. However the relevance analysis can give us more significant differences. The dendrogram, or grouping, does not depend on the scores of the models because it is made only of the distances between the descriptors. Even if there exists noise in the data, which may affect the scores of the model, we can expect that similar descriptors will give similar scores. The subgroup relevance analysis can illustrate how the distances, or the similarities, between the descriptors affect to the models. 

}

{Here, we { further} explain the advantage of the expression with the dendrogram. For example, we can easily choose \covrR\ if we do not want to use $Z_R$ if the importance is expressed as in Fig.~\ref{subgroupRelevanceAnalysis}. It enables us to find the next best route, that is, to go upward and try a new branch downward in the tree structure. We believe that this expression is much better than simply providing a list, and it is much easier to find out the operation-optimized {regression} models.

}

  We can conclude that the descriptor $C_R$ is strongly relevant when we define the subgroups at $d \sim 0$ and execute the leave-one-out experiment. The original relevance analysis is the special case of the subgroup relevance analysis. Therefore, the subgroup relevance analysis is a natural extension of the original relevance analysis.

  Here, we note the possible interpretation of the {regression} model in the context of condensed matter physics, where we know that physics should depend not on $J_4f$ but on $J_{4f}(1-g_J)$ in the effective model Hamiltonian. We, however, found more important descriptors, e.g., $Z_R$ and \covrR\ in the group R and $J_{3d}$ in the group T. It is more plausible that the regression model found {a relationship similar to the generalized Slater-Pauling curve for Curie temperature as a function of $C_R$ and $Z_T$ and $Z_R$}, and that the other effects are only marginal.\cite{SlaterPaulingTC} We introduced many descriptors that cannot appear in the atomic-scale effective model Hamiltonian, and the {regression} model simply selected the {{\it inter-scale} regression} model including the macro scale parameter $C_R$ {first and $Z_T$ and $Z_R$ next, which do not directly appear in the effective model Hamiltonian because their relationships are more apparent. It should be noted that the number of data, only about a hundred, is too few to discuss the details because it can easily change the prediction accuracy as discussed in the supporting information.\cite{supportinginfo} }

We cannot avoid errors in \TCs\ because of experimental errors and human errors. The latter is mainly because AtomWork does not allow web scraping. We examine the possibility of outlier detection using machine learning. We show a plot of experimental \TCs\ versus predicted ones in the supporting information.\cite{supportinginfo} The overall coincidence is good from 0K to $\sim$1300K, but there exist a few outliers. We mainly check the outliers of \TCs\ and fix the errors again and again if there are any. We found three major errors and a minor error. After fixing these errors, we evaluated the cross-validation test scores again for the best $n$ descriptors of the original {regression} model. The best $R^2$ was 0.96688. By using machine learning, it may be able possible to find data errors efficiently; however, it cannot detect data prediction of which appears consistent with the experimental values accidentally.

  We employed Pearson's correlation coefficient to define the distance in this study. However, there exist many choices for the distance. It depends on the problem whose representation is the most appropriate in the unsupervised learning part. We use the similarity, or distance, between materials to find the {regression} model, but usually discard the similarity between descriptors to make the {regression} model. We, however, utilized the latter similarity, and therefore took full advantage of the similarity of the data in this prescription.

  We showed that the distances between the descriptors are useful to illustrate the importance of descriptors and descriptor groups. This result is not strange when the descriptors have some physical meaning. There exists, however, minor discrepancies in the subgroup containing $Z_R$, $J_{3d}$, and $L_{3d}$ in the dendrogram. This is a limitation of this theory; however, it is possible to overcome this difficulty. We used the distance between the descriptors to explain the scores of the relevance analysis, but its inverse problem is also possible. We can set the value of distances between the descriptors, or the structures of the dendrogram, to be more consistent with the scores of the relevance analysis. 
  
  We can consider many variants of the subgroup relevance analysis. We took the best  {\it descriptor} from the subgroup shown in yellow in Fig.~\ref{SubgroupRelenvanceAnalysis}. Thus, we were able to show the best descriptors in the subgroup. Another method is to take the best {\it subgroup} in the downstream to a specified subgroup. Then, we will be able to understand the relationship among subgroups, and we can easily change them depending on the purpose. 

  Note that the Monte-Carlo tree search also utilizes the same nature of tree structures. There may be a route to find out the almost best {regression} model by utilizing subgroup decomposition without performing expensive exhaustive search. 

  In summary, we studied the data-driven approach on the Curie temperature of rare-earth transition metal stoichiometric alloys. We successfully made {regression} models that achieved high scores from our descriptors. We developed subgroup relevance analysis and successfully illustrated the importance, relationship, and structures among the descriptors from a huge list of exhaustive search. In addition, it shold be noted that our method makes full use of the similarity of the given data. 

\begin{acknowledgments}
  This work was partly supported by PRESTO and by the ``Materials Research by Information Integration'' Initiative (MI$^2$I) project of the Support Program for Starting Up Innovation Hub, both from the Japan Science and Technology Agency (JST), Japan; by the Elements Strategy Initiative Project under the auspices of MEXT; and also by MEXT as a social and scientific priority issue (Creation of New Functional Devices and High-Performance Materials to Support Next-Generation Industries; CDMSI) to be tackled by using a post-K computer. The calculations were partly carried out on Numerical Materials Simulator at NIMS. 

\end{acknowledgments}

\end{document}


\maketitle

\section{Descriptors}
We collected the experimental data of 101 binary compounds consisting of transition metals and rare-earth metals from the Atomwork database of NIMS \cite{atomwork}, including the crystal structure of the compounds and their observed {\TC}. To represent the structural and physical properties of each binary compound, we use a combination of 28 descriptors. We divide all 28 descriptors into three categories. 

The first category pertains to the descriptors describing the atomic properties of the transition-metal constituent, including the (1) atomic number ($Z_T$), (2) atomic radius ($r_T$), (3) covalent radius ($r^{cv}_{T}$), (4) ionization potential ($IP_{T}$), (5) electronegativity ($\chi_{T}$), (6) spin angular moment ($S_{3d}$), (7) orbital angular moment ($L_{3d}$), and (8) total angular moment ($J_{3d}$) of the 3$d$ electrons. The selection of these descriptors originates from the physical consideration that the intrinsic electronic and magnetic properties will determine the 3$d$ orbital splitting at transition-metal sites. 

In the same manner, we design the second category pertaining to the descriptors for describing the properties of the rare-earth metal constituent, including the (9) atomic number ($Z_R$), (10) atomic radius ($r_{R}$), (11) covalent radius ($r^{cv}_{R}$), (12) ionization potential ($IP_{R}$), (13) electronegativity ($\chi_{R}$), (14) spin angular moment ($S_{4f}$), (15) orbital angular moment ($L_{4f}$), and (16) total angular moment ($J_{4f}$) of the 4$f$ electrons. To capture the effect of the $4f$ electrons better, we add three additional descriptors for describing the properties of the constituent rare-earth metal ions, including (17) the Land$\acute{e}$ factor ($g_J$), (18) the projection of the total magnetic moment onto the total angular moment ($J_{4f}g_{J}$), and (19) the projection of the spin magnetic moment onto the total angular moment ($J_{4f}(1-g_J)$) of the 4$f$ electrons. The selection of these descriptors originates from the physical consideration that the magnitude of the magnetic moment will determine {\TC}. 

It has been well established that information related to the crystal structure is very valuable in relation to understanding the physics of binary compounds with transition metals and rare-earth metals. Therefore, we design the third category with structural descriptors that roughly represent the structural information at the transition metal and rare-earth metal sites, which are (20) the concentration of the transition metal ($C_T$), (21) the concentration of the rare-earth metal ($C_R$), (22) the average distance between a transition-metal site and the nearest transition-metal site ($d_{T-T}$), (23) the average distance between a transition-metal site and the nearest rare-earth-metal site ($d_{T-R}$), (24) the average distance between a rare-earth metal-site and the nearest rare-earth-metal site ($d_{R-R}$), (25) the average number of are-earth-metal sites surrounding a transition-metal site within  the distance less than 5.0 \AA ($N_{T-R}$), (26) the average number of rare-earth-metal sites surrounding a rare-earth-metal site within the distance less than 10.0 \AA ($N_{R-R}$), and (27) the average number of transition-metal sites surrounding a rare-earth-metal site within the distance less than 5.0 \AA ($N_{R-T}$). The values of these descriptors are calculated from the crystal structures of the compounds from the literature.

\section{Strong Relevance and Weak Relevance}
  We define the prediction ability $PA(\bm{S})$ of descriptors by the maximum prediction accuracy that the model can achieve by using the descriptors in a subset $\bm{s}$ of a set $\bm{S}$ of descriptors as follows:
\begin{equation}
PA(\bm{S}) = \max_{\forall \bm{s} \subset \bm{S}} R^2_s\text{,} 
\label{eq.PAS}
\end{equation}
where $R^2_s$ is the value of the coefficient of determination $R^2$ achieved by the model using a descriptor set $s$.
{$ ( R^2 = 1- \frac{\sum_i (y_i - y_i^{pred.})^2}{ \sum_i (y_i - \bar{y})^2 }$, where $y_i$, $y_i^{pred.}$, and $\bar y$ are the target value, the predicted value, and the man target value, respectively. )} 
On the basis of Eq.~(\ref{eq.PAS}), we can evaluate the relevance \cite{Relevanceanalysis1,Relevanceanalysis2} of a descriptor for the prediction of {\TC} by using the expected reduction in the prediction ability caused by removing this descriptor from the full set of descriptors. Let $\bm{D}$ be a full set of descriptors, $d_i$ a descriptor, and $\bm{D}_i = \bm{D} - \{d_i$\} the full set of descriptors after removing the descriptor $d_i$. The degree of relevance of the descriptors can be formalized as follows: 

  Strong relevance: a descriptor is strongly relevant if and only if 

\begin{equation}
PA(\bm{D})-PA(\bm{D}_i) = \max_{\forall s \subset \bm{D}} R^2_s - \max_{\forall s \subset \bm{D}_i} R^2_s > 0.
\label{eq.strong}
\end{equation}
Among the strongly relevant descriptors, a descriptor that causes a larger reduction in the prediction ability when it is removed can be considered as a strong one. The degree of relevance of a strongly relevant descriptor can be computationally estimated by using the {\bf leave-one-out} approach, i.e., by leaving out a descriptor in the currently considered descriptor set and testing how much the prediction accuracy is impaired.

  Weak relevance: a descriptor is weakly relevant if and only if 

\begin{align}
PA(\bm{D})-PA(\bm{D}_i) = \max_{\forall s \subset \bm{D}} R^2_s - \max_{\forall s \subset \bm{D}_i} R^2_s =0 \  {\bf and} \notag \\
\exists \bm{D}_i' \subset \bm{D}_i \   {\bf such \ that \ } PA(\{d_i,\bm{D}_i'\})-PA(\bm{D}_i') > 0.
\label{eq.weak}
\end{align}
It is clearly seen from Eq.~(\ref{eq.weak}) that estimation of the degree of relevance for the weakly relevant descriptors cannot be carried out in a straightforward manner as for the case of the strongly relevant descriptors. Weakly relevant descriptors are descriptors that are relevant for prediction, but they can be substituted by the other descriptors. We can only estimate the degree of relevance for this type of descriptor in specified contexts. For example, in terms of the prediction of {\TC}, the relevance of a descriptor for an atomic property of transition metal can be examined in the context that all of the descriptors for the atomic properties of rare-earth metals are included in the descriptor set. We define the following additional rule for comparing two weakly relevant descriptors: 

  Comparison between weakly relevant descriptors: A weakly relevant descriptor $d_i$ is said to be more relevant than the descriptor $d_j$ in the context of having a set of descriptors $\bm{M} (d_i, d_j \notin \bm{M})$ {\bf if and only if}

\begin{equation}
PA(\{d_i,\bm{M}\}) > PA(\{d_j,\bm{M}\}).
\label{eq.compare_weak}
\end{equation}

  A comparison of two weakly relevant descriptors can be computationally carried out by using the {\bf add-one-in} approach, i.e., by exclusively adding the two descriptors to the currently considered descriptor set and testing how much the prediction accuracy is improved.

\section{Best $R^2$ Scores and Descriptors}
We present a list of the best $R^2$ scores and descriptors in Table~\ref{nvsscoredescriptor}. It may appear that the difference in the scores is very small. We originally used ten times ten-fold cross validation (10$\times$10 CV). \cite{arXivDam2017} The best scores of the 10$\times$10 CV are the same for the two digits, i.e., they are 0.95X and 0.960 for $n$ = 5 to 10, where X varies.  Consequently, the plot of the scores versus $n$ shows a plateau.  We recognize that there exist non-negligible statistical errors which affects the relevance analysis. Next, we employ the leave-one-out cross validation because there exist no statistical errors and because we can obtain the most accurate scores from the data. Then, the best scores are the same for the three digits, i.e., they are 0.954X for $n$=5 to 8 in the leave-one-out cross validation, where there is a plateau in the score plot versus $n$. The difference between the scores becomes 10 times smaller in the latter.

\begin{table}
\caption{The number of descriptors vs the best $R^2$ score and descriptors.}
\label{nvsscoredescriptor}
\begin{tabular}{|p{0.5cm}|p{2cm}|p{10cm}|} 
\hline
n&score&descriptor(s) \\ \hline
1&0.3251{8}&\NRT\\
2&0.8701{5}& $C_R$, $Z_T$ \\
3&0.9422{2}& $C_R$, $Z_R$, $Z_T$ \\
4&0.9533{9}& $J_{3d}$, $C_R$, $Z_R$, $Z_T$ \\
5&0.9542{9}& $L_{3d}$, $J_{3d}$, $C_R$, $Z_R$, $Z_T$ \\
6&0.9543{9}& $L_{3d}$, $J_{3d}$, $\chi_T$, $C_R$, $Z_R$, $Z_T$ \\
7&0.9544{5}& $L_{3d}$, $J_{3d}$, $\chi_T$, $C_R$, $Z_R$, $Z_T$, \covrT \\
8&0.9544{5}& $L_{3d}$, $J_{3d}$, $\chi_T$, \IPT, $C_R$, $Z_R$, $Z_T$, \covrT \\
9&0.9535{1}& \dRT, $L_{3d}$, $J_{3d}$, $\chi_T$, \IPT, \covrR, $C_R$, \NRR, $J_{4f}g_J$ \\
10&0.9506{5}& \dRT, $L_{3d}$, $J_{3d}$, $\chi_T$, \IPT, \covrR, $C_R$, \NRR, $J_{4f}g_J$, $Z_T$ \\
11&0.9474{9}& \dRT, $L_{3d}$, $J_{3d}$, $\chi_T$, \IPT, $C_R$, \NRR, $Z_R$, $J_{4f}g_J$, $Z_T$, \covrT \\
12&0.9447{9}& \dRT, $L_{3d}$, $J_{3d}$, $\chi_T$, \IPT, \covrR, $C_R$, $C_T$, \NRR, $Z_R$, $J_{4f}g_J$, \covrT \\
13&0.9445{6}& $L_{3d}$, $J_{3d}$, $\chi_T$, $J_{4f}$, \IPT, $C_R$, $C_T$, \NRR, $Z_R$, $Z_T$, \NRT, \covrT, $L_{4f}$ \\
14&0.9432{2}& \dRT, $J_{4f}(1-g_J)$, $L_{3d}$, $J_{3d}$, $\chi_T$, $J_{4f}$, \IPR, \IPT, $C_R$, $C_T$, \NRR, $Z_R$, $Z_T$, \NRT \\
15&0.9424{5}& \dRT, $r_R$, $J_{4f}(1-g_J)$, $L_{3d}$, $J_{3d}$, $\chi_T$, \IPT, \covrR, $C_R$, $C_T$, \NRR, $Z_T$, \NRT, \covrT, $L_{4f}$ \\
16&0.9397{9}& \dRT, \dTT, $J_{4f}(1-g_J)$, $L_{3d}$, $J_{3d}$, $\chi_T$, $J_{4f}$, \IPR, \IPT, $C_R$, $C_T$, \NRR, $Z_R$, $Z_T$, \NRT, \covrT \\
17&0.9359{1}& \dRT, $r_R$, \dTT, $J_{4f}(1-g_J)$, $L_{3d}$, $J_{3d}$, $\chi_T$, $J_{4f}$, \IPR, \IPT, \covrR, $C_R$, $C_T$, \NRR, $Z_T$, \NRT, \covrT \\
18&0.9287{9}& \dRT, $r_R$, \dTT, $J_{4f}(1-g_J)$, $L_{3d}$, $J_{3d}$, $\chi_T$, \IPT, $C_R$, $C_T$, \dRR, $\chi_R$, \NRR, $Z_R$, $Z_T$, \covrT, $S_{4f}$, \NTR \\
19&0.9264{2}& \dRT, $r_R$, \dTT, $J_{4f}(1-g_J)$, $L_{3d}$, $J_{3d}$, $g_{J}$, $\chi_T$, \IPT, \covrR, $C_R$, $C_T$, \dRR, $\chi_R$, \NRR, $Z_R$, $Z_T$, \covrT, \NTR \\
\hline
\end{tabular}
\end{table}

\section{Prediction among the Best $n$ Models}
\begin{figure}
\includegraphics[width=15cm]{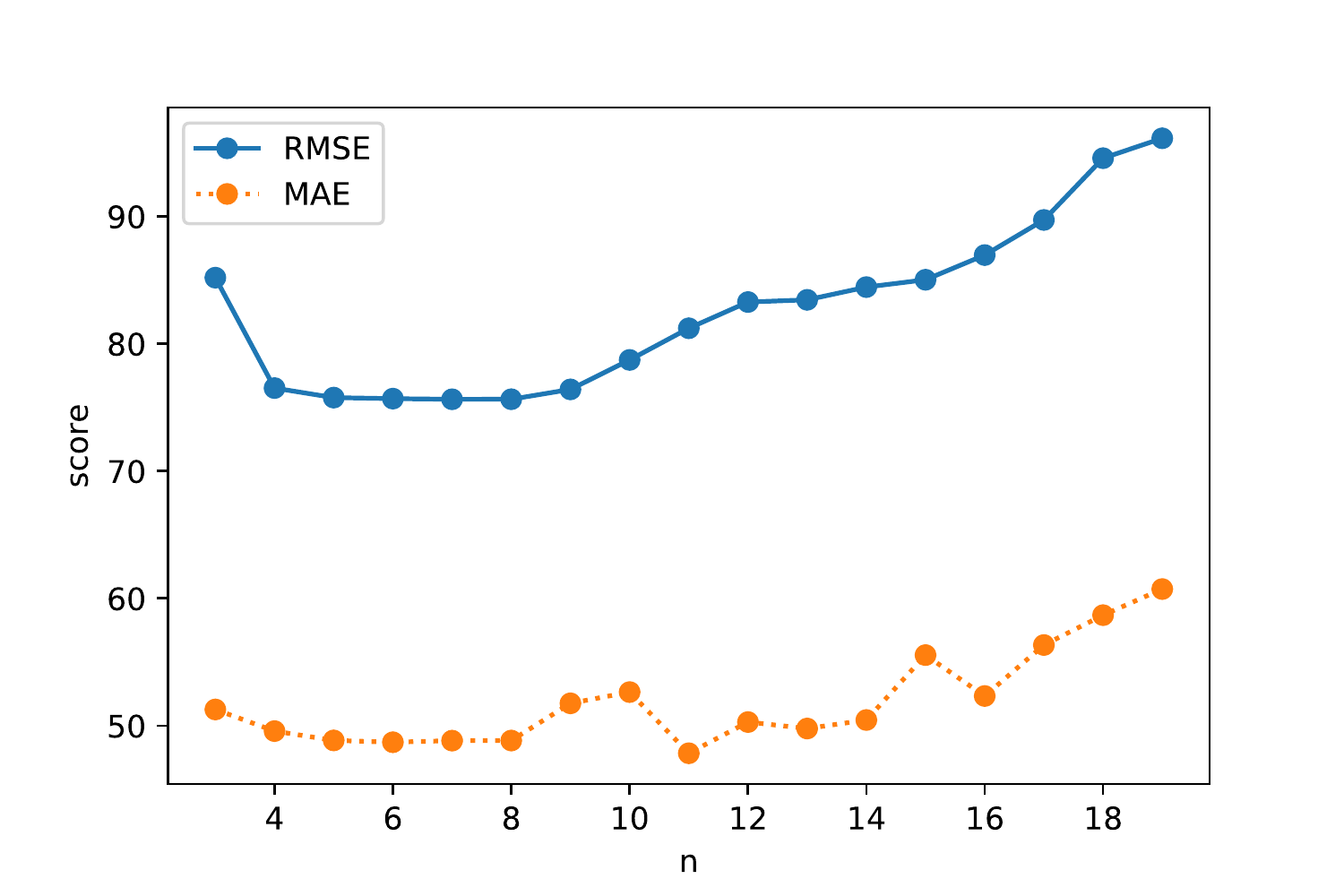}
\caption{ The best RMSE and MAE as functions of the number of descriptors ($^\circ$C).}
\label{nrmsemae}
\end{figure}

\begin{figure}
\includegraphics[width=15cm]{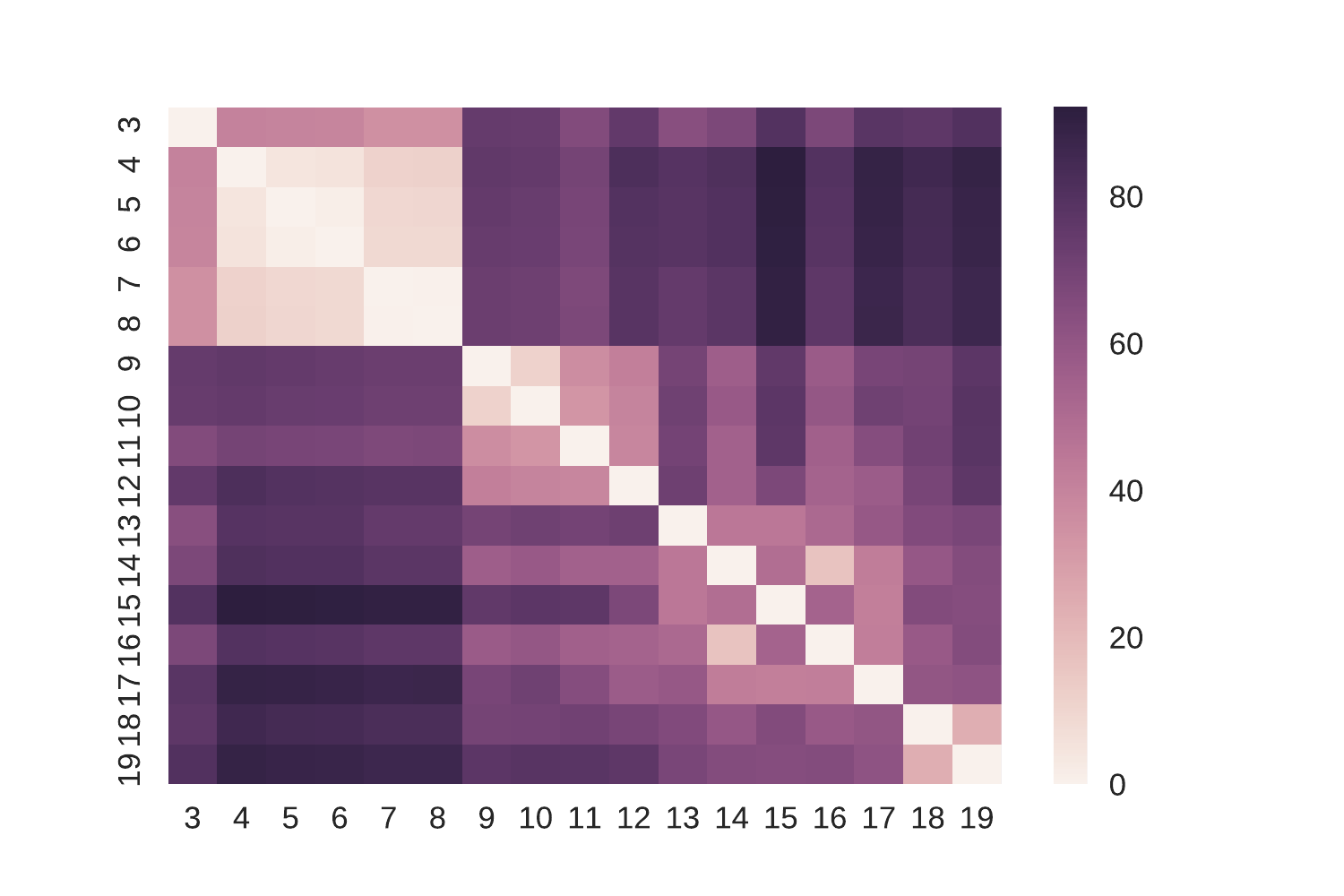}
\caption{Heatmap of "RMSE" between the best models with $n$ descriptors. }
\label{heatmaprmse}
\end{figure}

  We show the best scores of RMSE and MAE as a function of the number of descriptors ($n$) in the models in Fig.~\ref{nrmsemae}. The score changes gradually as a function of $n$. One may expect that their predictions are almost the same. 
  We also evaluate the "RMSE" between the leave-one-out cross validated test predictions of the best models with the $n$ descriptors in Fig.~\ref{heatmaprmse}. We can see that the predictions are almost the same for $n$=4 to 8; however, the deviations are larger in the other cases. Only the best models for $n$=4 to 8 give almost the same predictions. We can also see this trend from the kernel parameters.
  Note that these figures are the results before fixing the errors in the data. 

\section{Prediction among the Best $n$ Models after Fixing Errors}

\begin{figure}
\includegraphics[width=15cm]{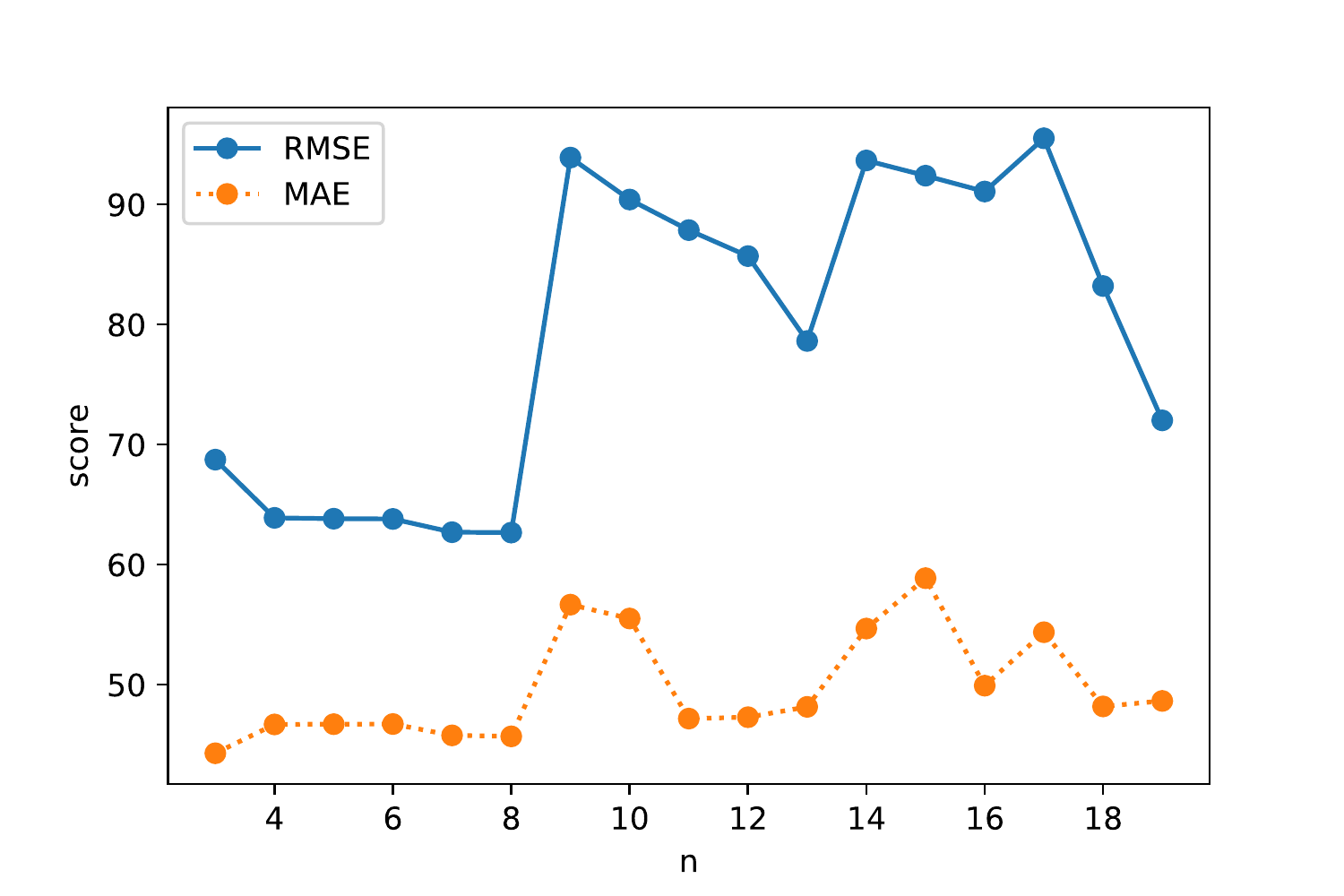}
\caption{ The best RMSE and MAE as functions of the number of descriptors ($^\circ$ C).}
\label{fix_nrmsemae}
\end{figure}

\begin{figure}
\includegraphics[width=15cm]{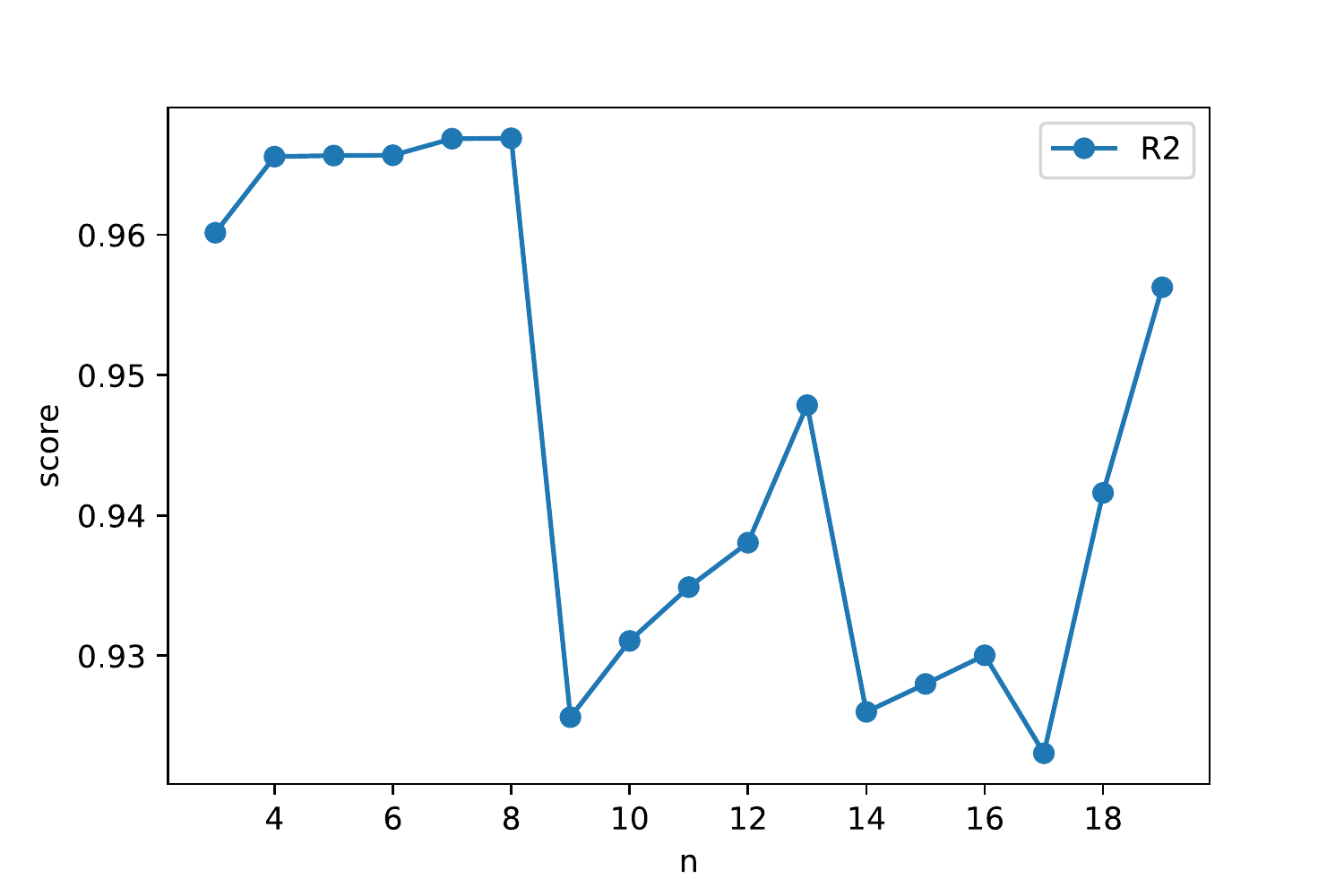}
\caption{ The best $R^2$ as a function of the number of descriptors.}
\label{fix_nr2}
\end{figure}

We show the scores for RMSE, MAE, and $R^2$ for the models in Table~\ref{nvsscoredescriptor} in Figs.~\ref{fix_nrmsemae} and \ref{fix_nr2}. { The models for $4\le n\le 8$ have high scores.}

\section{Experimental \TC\ versus CV-predicted \TC}

\begin{figure}
\includegraphics[width=15cm]{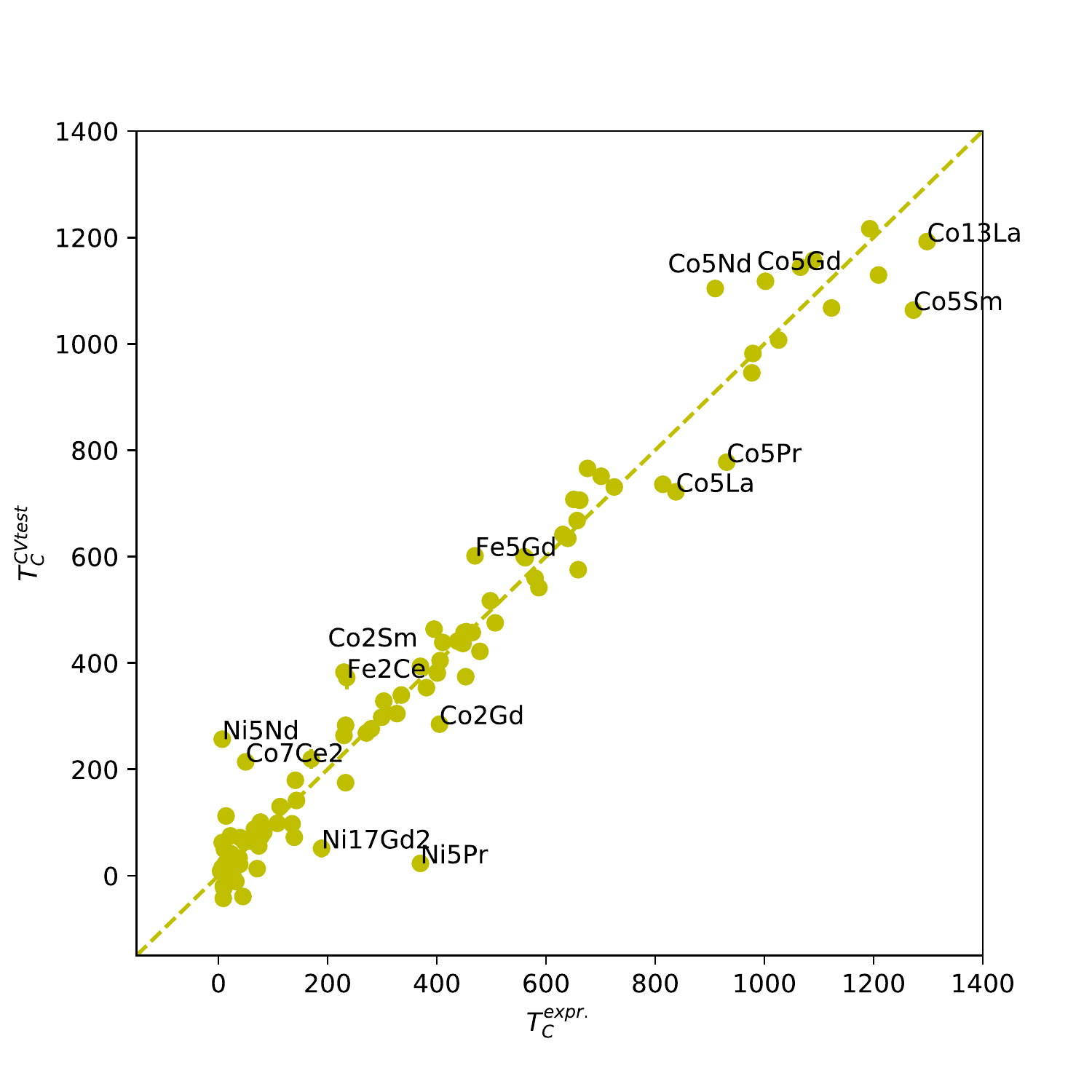}
\caption{Experimental \TC\ versus CV-predicted  \TC\ before fixing the errors.}
\label{TcexprvsTcpred}
\end{figure}

\begin{figure}
\includegraphics[width=15cm]{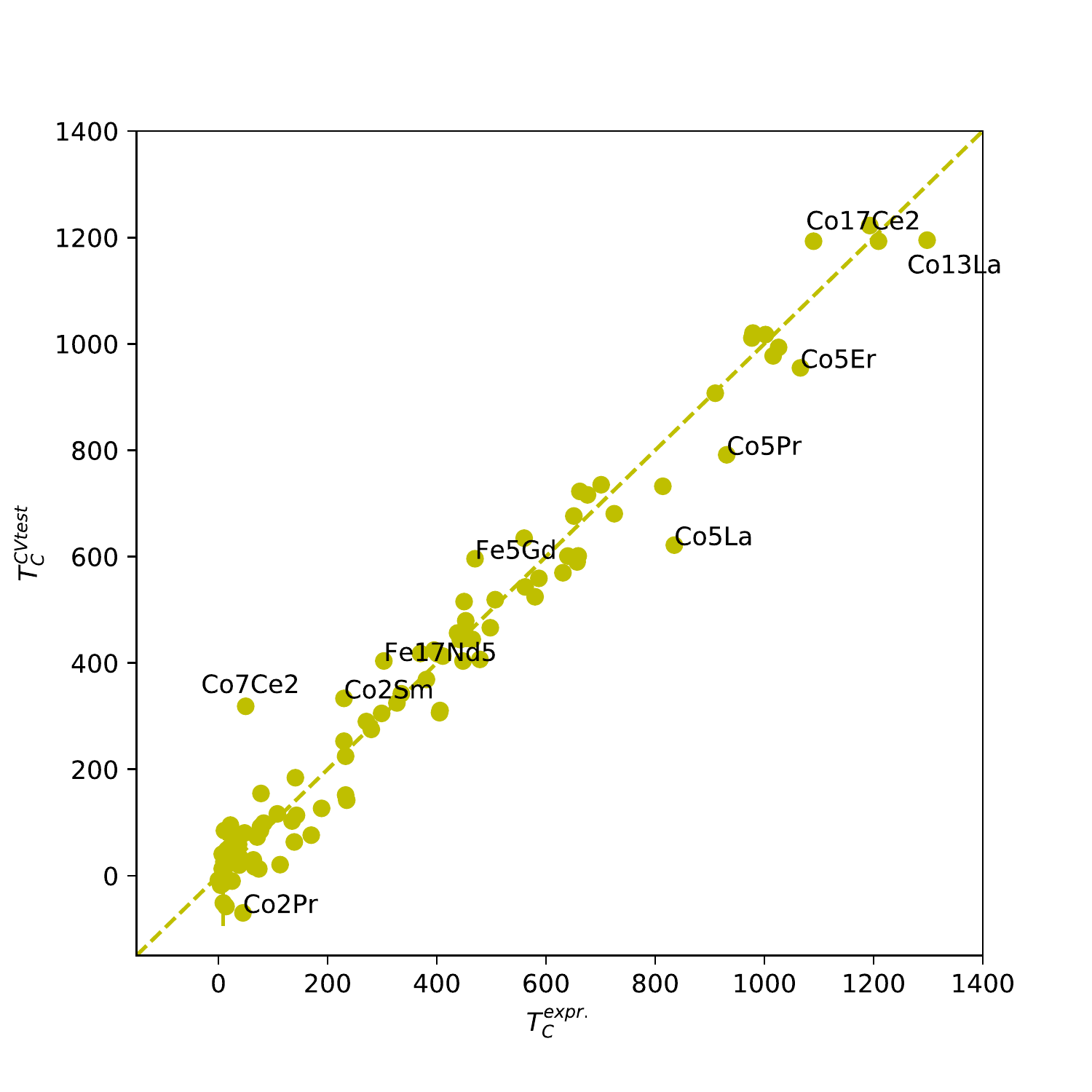}
\caption{Experimental \TC\ versus CV-predicted  \TC\ after fixing the errors.}
\label{fixed_TcexprvsTcpred}
\end{figure}

We plot the experimental \TC\ versus the CV-predicted \TC\ before and after fixing the errors in Fig.~\ref{TcexprvsTcpred} and \ref{fixed_TcexprvsTcpred}.  They show the mean and standard deviation of the predictions. The standard deviations are shown as bars, but almost all of them are smaller than the sizes of the symbols. 

The overall coincidence is good from 0K to approximately 1300K, but we find a few outliers in Fig.~\ref{TcexprvsTcpred}. For example, the experimental \TCs\ of SmCo$_5$ and PrNi$_5$ are much higher than the predicted ones, whereas the experimental \TCs\ are much smaller for NdCo$_5$ and NdNi$_5$. We find three major errors and a minor error in the experimental \TCs\ including those for SmCo$_5$ and PrNi$_5$. 

A new plot obtained after fixing the errors is shown in Fig.\ref{fixed_TcexprvsTcpred}. The predicted values of NdCo$_5$ and NdNi$_5$ now are almost the same as the experimental values. We find other outliers, such as the data for Ce$_2$Co$_7$ and RCo$_5$. However, it appears that these are not because of the errors in the data.

\section{Predicted \TCs\ for (RE)Fe$_{12}$}

  We examine the prediction ability of the best regression model. We apply the best regression model to (RE)Fe$_{12}$, which was recently synthesized and attracts much attention. The existing experimental \TCs\ are 508K for NdFe$_{12}$, \cite{HirayamaPrivate} 586K for SmFe$_{12}$, \cite{Hirayama2017} and 483K for YFe$_{12}$. \cite{Suzuki2017}. On the other hand, the corresponding predicted \TCs\ are 490(19)K, 581(15)K, and 396(10)K, where the crystal structures are obtained from the first-principles calculation and we substituted the Z and quantum-number-related descriptors of La for those of Y.\cite{ReFe12Harashima}  The coincidences of the values of NdFe$_{12}$ and SmFe$_{12}$ are fairly good considering the fact that we do not have the structure data in the training set. The predicted values for DyFe$_{12}$ and GdFe$_{12}$ are 470(11)K and 600(13)K, respectively. {However, these predicted values decrease by 120--180K after fixing the errors in the data. The predicted values depend on the value of the L2 penalty term. We add {this information} as reference. }

\section{List of Descriptors}

We list the original descriptors and \TCs\ before fixing the errors in Tables~\ref{c00}-\ref{c22}.  
We list the final descriptors and \TCs\ after fixing the errors in Tables~\ref{content00}-\ref{content22}.
The number of original materials was 101, but we found a non-stoichiometry material, which was deleted.

\begin{table}
\begin{tabular}{cccccccccc}
material&$T_C$&$Z_R$&$r_R$&$r_R^{cov}$&$IP_R$&$\chi_R$&$S_{4f}$&$L_{4f}$&$J_{4f}$\\
Co17Ce2(Zn17Th2)&1090&58&181&204&534.4&1.12&0.5&3&2.5\\
Co5Ce(CaCu5)&662&58&181&204&534.4&1.12&0.5&3&2.5\\
Co7Ce2(Gd2Co7)&50&58&181&204&534.4&1.12&0.5&3&2.5\\
Fe17Ce2(Zn17Th2)&233&58&181&204&534.4&1.12&0.5&3&2.5\\
Fe2Ce(MgCu2)&235&58&181&204&534.4&1.12&0.5&3&2.5\\
Co2Dy(MgCu2)&141&66&178&192&573.0&1.22&2.5&5&7.5\\
Co3Dy(PuNi3)&450&66&178&192&573.0&1.22&2.5&5&7.5\\
Co5Dy(CaCu5)&977&66&178&192&573.0&1.22&2.5&5&7.5\\
Mn23Dy6(Th6Mn23)&443&66&178&192&573.0&1.22&2.5&5&7.5\\
Mn2Dy(MgZn2)&37&66&178&192&573.0&1.22&2.5&5&7.5\\
Ni17Dy2(Th2Ni17)&170&66&178&192&573.0&1.22&2.5&5&7.5\\
Ni2Dy(MgCu2)&28&66&178&192&573.0&1.22&2.5&5&7.5\\
Ni5Dy(CaCu5)&13&66&178&192&573.0&1.22&2.5&5&7.5\\
NiDy(FeB-b)&64&66&178&192&573.0&1.22&2.5&5&7.5\\
Co2Er(TbFe2)&39&68&176&189&589.3&1.24&1.5&6&7.5\\
Co3Er(PuNi3)&401&68&176&189&589.3&1.24&1.5&6&7.5\\
Co5Er(CaCu5)&1066&68&176&189&589.3&1.24&1.5&6&7.5\\
Co7Er(Cu5.44Tb0.78)&1123&68&176&189&589.3&1.24&1.5&6&7.5\\
CoEr3(Fe3C)&7&68&176&189&589.3&1.24&1.5&6&7.5\\
Fe17Er2(Th2Ni17)&299&68&176&189&589.3&1.24&1.5&6&7.5\\
Fe23Er6(Th6Mn23)&498&68&176&189&589.3&1.24&1.5&6&7.5\\
Fe2Er(MgCu2)&587&68&176&189&589.3&1.24&1.5&6&7.5\\
Ni3Er(PuNi3)&66&68&176&189&589.3&1.24&1.5&6&7.5\\
Ni5Er(CaCu5)&11&68&176&189&589.3&1.24&1.5&6&7.5\\
Ni7Er2(Gd2Co7)&74&68&176&189&589.3&1.24&1.5&6&7.5\\
NiEr(FeB-b)&12&68&176&189&589.3&1.24&1.5&6&7.5\\
Ni2Eu(MgCu2)&139&63&180&198&547.1&1.2&3.0&3&0.0\\
Co17Gd2(Zn17Th2)&1209&64&180&196&593.4&1.2&3.5&0&3.5\\
Co2Gd(MgCu2)&405&64&180&196&593.4&1.2&3.5&0&3.5\\
Co3Gd(PuNi3)&631&64&180&196&593.4&1.2&3.5&0&3.5\\
Co3Gd4(Ho6Co4.5)&233&64&180&196&593.4&1.2&3.5&0&3.5\\
Co5Gd(CaCu5)&1002&64&180&196&593.4&1.2&3.5&0&3.5\\
CoGd3(Fe3C)&143&64&180&196&593.4&1.2&3.5&0&3.5\\
Fe17Gd2(Zn17Th2)&479&64&180&196&593.4&1.2&3.5&0&3.5\\
Fe23Gd6(Th6Mn23)&659&64&180&196&593.4&1.2&3.5&0&3.5\\
Fe2Gd(MgCu2)&814&64&180&196&593.4&1.2&3.5&0&3.5\\
Fe3Gd(PuNi3)&725&64&180&196&593.4&1.2&3.5&0&3.5\\
Fe5Gd(CaCu5)&470&64&180&196&593.4&1.2&3.5&0&3.5\\
Mn23Gd6(Th6Mn23)&465&64&180&196&593.4&1.2&3.5&0&3.5\\
Mn2Gd(MgCu2)&135&64&180&196&593.4&1.2&3.5&0&3.5\\
\end{tabular}
\caption{Descriptors from the 1st to the 40th material.}
\label{c00}
\end{table}

\begin{table}
\begin{tabular}{cccccccccc}
$g_J$&$J_{4f} g_J$&$J_{4f} (g_J-1)$&$Z_T$&$r_T$&$r_T^{cov}$&$IP_T$&$\chi_T$&$S_3d$&$L_3d$\\
0.8571&2.14275&-0.35725&27.0&125.0&150.0&760.4&1.88&1.5&3.0\\
0.8571&2.14275&-0.35725&27.0&125.0&150.0&760.4&1.88&1.5&3.0\\
0.8571&2.14275&-0.35725&27.0&125.0&150.0&760.4&1.88&1.5&3.0\\
0.8571&2.14275&-0.35725&26.0&126.0&152.0&762.5&1.83&2.0&2.0\\
0.8571&2.14275&-0.35725&26.0&126.0&152.0&762.5&1.83&2.0&2.0\\
1.3333&9.99975&2.49975&27.0&125.0&150.0&760.4&1.88&1.5&3.0\\
1.3333&9.99975&2.49975&27.0&125.0&150.0&760.4&1.88&1.5&3.0\\
1.3333&9.99975&2.49975&27.0&125.0&150.0&760.4&1.88&1.5&3.0\\
1.3333&9.99975&2.49975&25.0&127.0&161.0&717.3&1.55&2.5&0.0\\
1.3333&9.99975&2.49975&25.0&127.0&161.0&717.3&1.55&2.5&0.0\\
1.3333&9.99975&2.49975&28.0&124.0&124.0&737.1&1.91&1.0&3.0\\
1.3333&9.99975&2.49975&28.0&124.0&124.0&737.1&1.91&1.0&3.0\\
1.3333&9.99975&2.49975&28.0&124.0&124.0&737.1&1.91&1.0&3.0\\
1.3333&9.99975&2.49975&28.0&124.0&124.0&737.1&1.91&1.0&3.0\\
1.2&9.0&1.5&27.0&125.0&150.0&760.4&1.88&1.5&3.0\\
1.2&9.0&1.5&27.0&125.0&150.0&760.4&1.88&1.5&3.0\\
1.2&9.0&1.5&27.0&125.0&150.0&760.4&1.88&1.5&3.0\\
1.2&9.0&1.5&27.0&125.0&150.0&760.4&1.88&1.5&3.0\\
1.2&9.0&1.5&27.0&125.0&150.0&760.4&1.88&1.5&3.0\\
1.2&9.0&1.5&26.0&126.0&152.0&762.5&1.83&2.0&2.0\\
1.2&9.0&1.5&26.0&126.0&152.0&762.5&1.83&2.0&2.0\\
1.2&9.0&1.5&26.0&126.0&152.0&762.5&1.83&2.0&2.0\\
1.2&9.0&1.5&28.0&124.0&124.0&737.1&1.91&1.0&3.0\\
1.2&9.0&1.5&28.0&124.0&124.0&737.1&1.91&1.0&3.0\\
1.2&9.0&1.5&28.0&124.0&124.0&737.1&1.91&1.0&3.0\\
1.2&9.0&1.5&28.0&124.0&124.0&737.1&1.91&1.0&3.0\\
0.0&0.0&0.0&28.0&124.0&124.0&737.1&1.91&1.0&3.0\\
2.0&7.0&3.5&27.0&125.0&150.0&760.4&1.88&1.5&3.0\\
2.0&7.0&3.5&27.0&125.0&150.0&760.4&1.88&1.5&3.0\\
2.0&7.0&3.5&27.0&125.0&150.0&760.4&1.88&1.5&3.0\\
2.0&7.0&3.5&27.0&125.0&150.0&760.4&1.88&1.5&3.0\\
2.0&7.0&3.5&27.0&125.0&150.0&760.4&1.88&1.5&3.0\\
2.0&7.0&3.5&27.0&125.0&150.0&760.4&1.88&1.5&3.0\\
2.0&7.0&3.5&26.0&126.0&152.0&762.5&1.83&2.0&2.0\\
2.0&7.0&3.5&26.0&126.0&152.0&762.5&1.83&2.0&2.0\\
2.0&7.0&3.5&26.0&126.0&152.0&762.5&1.83&2.0&2.0\\
2.0&7.0&3.5&26.0&126.0&152.0&762.5&1.83&2.0&2.0\\
2.0&7.0&3.5&26.0&126.0&152.0&762.5&1.83&2.0&2.0\\
2.0&7.0&3.5&25.0&127.0&161.0&717.3&1.55&2.5&0.0\\
2.0&7.0&3.5&25.0&127.0&161.0&717.3&1.55&2.5&0.0\\
\end{tabular}
\caption{Descriptors from the 1st to the 40th material. (cont.)}
\label{c01}
\end{table}

\begin{table}
\begin{tabular}{cccccccc}
$J_3d$&$C_T$&$C_R$&$d_{T-T}$&$d_{R-T}$&$d_{R-R}$&$N_{T-R}$&$N_{R-R}$\\
4.5&0.06846&0.00805&2.37126&2.79635&4.07441&5.76471&38.0\\
4.5&0.05917&0.01183&2.46168&2.84518&4.01800&8.4&58.0\\
4.5&0.05449&0.01557&2.45226&2.83193&3.12910&10.0&68.5\\
4.0&0.06572&0.00773&2.40832&2.83201&4.13808&3.64706&38.0\\
4.0&0.04110&0.02055&2.58165&3.02725&3.16186&14.0&86.0\\
4.5&0.04294&0.02147&2.54417&2.98330&3.11596&14.0&86.0\\
4.5&0.05123&0.01708&2.47726&2.87697&3.13757&10.88889&78.0\\
4.5&0.08345&0.01192&1.54703&1.22008&3.98720&6.57143&58.0\\
2.5&0.04780&0.01247&2.53316&3.05688&3.57105&5.47826&50.0\\
2.5&0.03659&0.0183&2.68347&3.14665&3.28657&14.0&86.0\\
4.0&0.07398&0.0087&2.25952&2.60907&4.33000&4.35294&36.0\\
4.0&0.04366&0.02183&2.53003&2.96672&3.09864&14.0&86.0\\
4.0&0.06136&0.01227&2.43184&2.81112&3.96900&8.4&58.0\\
4.0&0.02502&0.02502&2.45561&2.81558&3.55253&11.0&106.0\\
4.5&0.04364&0.02182&2.53038&2.96714&3.09907&14.0&86.0\\
4.5&0.05188&0.01729&2.46788&2.86461&3.12373&10.88889&78.0\\
4.5&0.06033&0.01207&2.44450&2.82267&4.00400&8.4&58.0\\
4.5&0.08541&0.0122&1.56674&1.23563&4.03800&8.28571&58.0\\
4.5&0.01019&0.03056&4.19439&2.68390&3.32898&16.0&132.0\\
4.0&0.06635&0.00781&2.32148&2.81572&4.14550&3.64706&36.0\\
4.0&0.05332&0.01391&2.44254&2.94752&3.44330&8.6087&62.0\\
4.0&0.04162&0.02081&2.57069&3.01440&3.14844&14.0&86.0\\
4.0&0.05241&0.01747&2.46367&2.85376&3.11046&10.88889&78.0\\
4.0&0.06173&0.01235&2.42800&2.80361&3.96600&8.4&58.0\\
4.0&0.05579&0.01594&2.42823&2.81128&3.10758&10.0&68.5\\
4.0&0.02567&0.02567&2.43010&2.79557&3.52136&15.0&106.0\\
4.0&0.04056&0.02028&2.59296&3.04052&3.17572&14.0&86.0\\
4.5&0.06896&0.00811&2.36486&2.79001&4.06341&5.76471&38.0\\
4.5&0.04149&0.02074&2.57352&3.01771&3.15190&14.0&86.0\\
4.5&0.05047&0.01682&2.48879&2.89171&3.15397&10.22222&66.0\\
4.5&0.02116&0.02539&2.02400&2.85960&3.40595&12.6&107.0\\
4.5&0.05900&0.0118&2.44924&2.86539&3.97300&8.4&58.0\\
4.5&0.00941&0.02822&4.28765&2.74391&3.45177&10.0&122.0\\
4.0&0.06518&0.00767&2.42005&2.96247&4.17250&4.35294&36.0\\
4.0&0.05210&0.01359&2.46148&2.97038&3.47000&8.6087&62.0\\
4.0&0.03942&0.01971&2.61771&3.06954&3.20603&14.0&86.0\\
4.0&0.04746&0.01582&2.52733&2.95576&3.22844&10.22222&66.0\\
4.0&0.05992&0.01198&2.41500&2.78860&4.13000&8.4&58.0\\
2.5&0.04705&0.01227&2.54660&3.07310&3.59000&5.47826&50.0\\
2.5&0.03451&0.01725&2.73650&3.20883&3.35152&6.0&70.0\\
\end{tabular}
\caption{Descriptors from the 1st to the 40th material. ({\it cont. 2})}
\label{c02}
\end{table}

\begin{table}
\begin{tabular}{cccccccccc}
material&$T_C$&$Z_R$&$r_R$&$r_R^{cov}$&$IP_R$&$\chi_R$&$S_{4f}$&$L_{4f}$&$J_{4f}$\\
NI17Gd2(Th2Ni17)&189&64&180&196&593.4&1.2&3.5&0&3.5\\
Ni2Gd(MgCu2)&77&64&180&196&593.4&1.2&3.5&0&3.5\\
Ni3Gd(PuNi3)&113&64&180&196&593.4&1.2&3.5&0&3.5\\
Ni5Gd(CaCu5)&32&64&180&196&593.4&1.2&3.5&0&3.5\\
NiGd(TlI)&71&64&180&196&593.4&1.2&3.5&0&3.5\\
Co2Ho(MgCu2)&83&67&176&192&581.0&1.23&2.0&6&8.0\\
Co5Ho(CaCu5)&1026&67&176&192&581.0&1.23&2.0&6&8.0\\
CoHo3(Fe3C)&10&67&176&192&581.0&1.23&2.0&6&8.0\\
Fe17Ho2(Th2Ni17)&335&67&176&192&581.0&1.23&2.0&6&8.0\\
Fe23Ho6(Th6Mn23)&507&67&176&192&581.0&1.23&2.0&6&8.0\\
Fe2Ho(MgCu2)&560&67&176&192&581.0&1.23&2.0&6&8.0\\
Mn2Ho(MgCu2)&25&67&176&192&581.0&1.23&2.0&6&8.0\\
Ni2Ho(MgCu2)&16&67&176&192&581.0&1.23&2.0&6&8.0\\
Ni5Ho(CaCu5)&14&67&176&192&581.0&1.23&2.0&6&8.0\\
NiHo(FeB-b)&38&67&176&192&581.0&1.23&2.0&6&8.0\\
Co13La(NaZn13)&1298&57&187&207&538.1&1.1&0.0&0&0.0\\
Co5La(CaCu5)&838&57&187&207&538.1&1.1&0.0&0&0.0\\
Fe2Lu(MgCu2)&580&71&174&187&523.5&1.27&0.0&0&0.0\\
Co2Nd(MgCu2)&108&60&181&201&533.1&1.14&1.5&6&4.5\\
Co3Nd(PuNi3)&381&60&181&201&533.1&1.14&1.5&6&4.5\\
Co5Nd(CaCu5)&910&60&181&201&533.1&1.14&1.5&6&4.5\\
CoNd3(Fe3C)&27&60&181&201&533.1&1.14&1.5&6&4.5\\
Fe17Nd2(Zn17Th2)&327&60&181&201&533.1&1.14&1.5&6&4.5\\
Fe17Nd5(Nd5Fe17)&303&60&181&201&533.1&1.14&1.5&6&4.5\\
Fe2Nd(MgCu2)&453&60&181&201&533.1&1.14&1.5&6&4.5\\
Mn23Nd6(Th6Mn23)&438&60&181&201&533.1&1.14&1.5&6&4.5\\
Ni2Nd(MgCu2)&9&60&181&201&533.1&1.14&1.5&6&4.5\\
Ni5Nd(CaCu5)&7&60&181&201&533.1&1.14&1.5&6&4.5\\
Co2Pr(MgCu2)&45&59&182&203&527.0&1.13&1.0&5&4.0\\
Co5Pr(CaCu5)&931&59&182&203&527.0&1.13&1.0&5&4.0\\
CoPr3(Fe3C)&14&59&182&203&527.0&1.13&1.0&5&4.0\\
Fe17Pr2(Zn17Th2)&280&59&182&203&527.0&1.13&1.0&5&4.0\\
Mn23Pr6(Th6Mn23)&448&59&182&203&527.0&1.13&1.0&5&4.0\\
Ni5Pr(CaCu5)&370&59&182&203&527.0&1.13&1.0&5&4.0\\
Co17Sm2(Zn17Th2)&1193&62&180&198&544.5&1.17&2.5&5&2.5\\
Co2Sm(MgCu2)&230&62&180&198&544.5&1.17&2.5&5&2.5\\
Co5Sm(CaCu5)&1273&62&180&198&544.5&1.17&2.5&5&2.5\\
CoSm3(Fe3C)&78&62&180&198&544.5&1.17&2.5&5&2.5\\
Fe17Sm2(Zn17Th2)&395&62&180&198&544.5&1.17&2.5&5&2.5\\
Fe2Sm(MgCu2)&676&62&180&198&544.5&1.17&2.5&5&2.5\\
\end{tabular}
\caption{Descriptors from the 41st to the 80th material.}
\label{c10}
\end{table}

\begin{table}
\begin{tabular}{cccccccccc}
$g_J$&$J_{4f} g_J$&$J_{4f} (g_J-1)$&$Z_T$&$r_T$&$r_T^{cov}$&$IP_T$&$\chi_T$&$S_3d$&$L_3d$\\
2.0&7.0&3.5&28.0&124.0&124.0&737.1&1.91&1.0&3.0\\
2.0&7.0&3.5&28.0&124.0&124.0&737.1&1.91&1.0&3.0\\
2.0&7.0&3.5&28.0&124.0&124.0&737.1&1.91&1.0&3.0\\
2.0&7.0&3.5&28.0&124.0&124.0&737.1&1.91&1.0&3.0\\
2.0&7.0&3.5&28.0&124.0&124.0&737.1&1.91&1.0&3.0\\
1.25&10.0&2.0&27.0&125.0&150.0&760.4&1.88&1.5&3.0\\
1.25&10.0&2.0&27.0&125.0&150.0&760.4&1.88&1.5&3.0\\
1.25&10.0&2.0&27.0&125.0&150.0&760.4&1.88&1.5&3.0\\
1.25&10.0&2.0&26.0&126.0&152.0&762.5&1.83&2.0&2.0\\
1.25&10.0&2.0&26.0&126.0&152.0&762.5&1.83&2.0&2.0\\
1.25&10.0&2.0&26.0&126.0&152.0&762.5&1.83&2.0&2.0\\
1.25&10.0&2.0&25.0&127.0&161.0&717.3&1.55&2.5&0.0\\
1.25&10.0&2.0&28.0&124.0&124.0&737.1&1.91&1.0&3.0\\
1.25&10.0&2.0&28.0&124.0&124.0&737.1&1.91&1.0&3.0\\
1.25&10.0&2.0&28.0&124.0&124.0&737.1&1.91&1.0&3.0\\
0.0&0.0&0.0&27.0&125.0&150.0&760.4&1.88&1.5&3.0\\
0.0&0.0&0.0&27.0&125.0&150.0&760.4&1.88&1.5&3.0\\
0.0&0.0&0.0&26.0&126.0&152.0&762.5&1.83&2.0&2.0\\
0.7273&3.27285&-1.22715&27.0&125.0&150.0&760.4&1.88&1.5&3.0\\
0.7273&3.27285&-1.22715&27.0&125.0&150.0&760.4&1.88&1.5&3.0\\
0.7273&3.27285&-1.22715&27.0&125.0&150.0&760.4&1.88&1.5&3.0\\
0.7273&3.27285&-1.22715&27.0&125.0&150.0&760.4&1.88&1.5&3.0\\
0.7273&3.27285&-1.22715&26.0&126.0&152.0&762.5&1.83&2.0&2.0\\
0.7273&3.27285&-1.22715&26.0&126.0&152.0&762.5&1.83&2.0&2.0\\
0.7273&3.27285&-1.22715&26.0&126.0&152.0&762.5&1.83&2.0&2.0\\
0.7273&3.27285&-1.22715&25.0&127.0&161.0&717.3&1.55&2.5&0.0\\
0.7273&3.27285&-1.22715&28.0&124.0&124.0&737.1&1.91&1.0&3.0\\
0.7273&3.27285&-1.22715&28.0&124.0&124.0&737.1&1.91&1.0&3.0\\
0.8&3.2&-0.8&27.0&125.0&150.0&760.4&1.88&1.5&3.0\\
0.8&3.2&-0.8&27.0&125.0&150.0&760.4&1.88&1.5&3.0\\
0.8&3.2&-0.8&27.0&125.0&150.0&760.4&1.88&1.5&3.0\\
0.8&3.2&-0.8&26.0&126.0&152.0&762.5&1.83&2.0&2.0\\
0.8&3.2&-0.8&25.0&127.0&161.0&717.3&1.55&2.5&0.0\\
0.8&3.2&-0.8&28.0&124.0&124.0&737.1&1.91&1.0&3.0\\
0.2857&0.71425&-1.78575&27.0&125.0&150.0&760.4&1.88&1.5&3.0\\
0.2857&0.71425&-1.78575&27.0&125.0&150.0&760.4&1.88&1.5&3.0\\
0.2857&0.71425&-1.78575&27.0&125.0&150.0&760.4&1.88&1.5&3.0\\
0.2857&0.71425&-1.78575&27.0&125.0&150.0&760.4&1.88&1.5&3.0\\
0.2857&0.71425&-1.78575&26.0&126.0&152.0&762.5&1.83&2.0&2.0\\
0.2857&0.71425&-1.78575&26.0&126.0&152.0&762.5&1.83&2.0&2.0\\
\end{tabular}
\caption{Descriptors from the 41st to the 80th material. (cont.)}
\label{c11}
\end{table}

\begin{table}
\begin{tabular}{cccccccc}
$J_3d$&$C_T$&$C_R$&$d_{T-T}$&$d_{R-T}$&$d_{R-R}$&$N_{T-R}$&$N_{R-R}$\\
4.0&0.06915&0.00814&2.36257&2.72806&4.23700&4.35294&36.0\\
4.0&0.04265&0.02133&2.54983&2.98994&3.12289&14.0&86.0\\
4.0&0.05057&0.01686&2.49359&2.88764&3.14721&10.22222&66.0\\
4.0&0.06042&0.01208&2.43690&2.83421&3.96500&8.4&58.0\\
4.0&0.02420&0.0242&2.58495&2.89772&3.58845&11.0&92.0\\
4.5&0.04333&0.02167&2.53639&2.97418&3.10643&14.0&86.0\\
4.5&0.05994&0.01199&2.44444&2.84056&3.97900&8.4&58.0\\
4.5&0.01001&0.03003&4.20705&2.69174&3.36592&14.0&131.33333\\
4.0&0.06652&0.00783&2.32484&2.81005&4.15150&3.64706&36.0\\
4.0&0.05259&0.01372&2.45374&2.96104&3.45909&8.6087&62.0\\
4.0&0.04058&0.02029&2.59261&3.04010&3.17528&14.0&86.0\\
2.5&0.03852&0.01926&2.63800&3.09333&3.23088&14.0&86.0\\
4.0&0.04410&0.02205&2.52154&2.95677&3.08825&14.0&86.0\\
4.0&0.06135&0.01227&2.43006&2.81343&3.96300&8.4&58.0\\
4.0&0.02533&0.02533&2.44355&2.80573&3.53764&11.0&106.0\\
4.5&0.07100&0.00546&2.37741&3.29960&5.67850&2.46154&26.0\\
4.5&0.05567&0.01113&2.47327&2.95084&3.97000&8.4&52.0\\
4.0&0.04234&0.02117&2.55619&2.99740&3.13068&14.0&86.0\\
4.5&0.04096&0.02048&2.58448&3.03057&3.16532&14.0&86.0\\
4.5&0.04914&0.01638&2.51634&2.91590&3.17848&10.22222&66.0\\
4.5&0.05728&0.01146&2.46505&2.90407&3.98400&8.4&52.0\\
4.5&0.00897&0.02692&4.33408&2.77265&3.53652&10.0&115.33333\\
4.0&0.06421&0.00755&2.39377&3.07380&3.91324&4.35294&38.0\\
4.0&0.04675&0.01375&2.33911&2.96841&3.25413&7.52941&56.2\\
4.0&0.03854&0.01927&2.63751&3.09275&3.23027&14.0&86.0\\
2.5&0.04510&0.01177&2.58265&3.11660&3.64082&5.47826&50.0\\
4.0&0.04167&0.02083&2.56973&3.01328&3.14727&14.0&86.0\\
4.0&0.05919&0.01184&2.44789&2.86077&3.97300&8.4&58.0\\
4.5&0.04093&0.02046&2.58518&3.03140&3.16619&14.0&86.0\\
4.5&0.05774&0.01155&2.46101&2.89310&3.98200&8.4&52.0\\
4.5&0.00890&0.0267&4.33876&2.77540&3.55680&10.0&115.33333\\
4.0&0.06417&0.00755&2.41782&2.86035&4.15442&3.64706&38.0\\
2.5&0.04465&0.01165&2.59140&3.12717&3.65316&5.47826&50.0\\
4.0&0.05914&0.01183&2.44823&2.86193&3.97300&8.4&58.0\\
4.5&0.06835&0.00804&2.36874&2.80001&4.07008&5.76471&38.0\\
4.5&0.04180&0.0209&2.56715&3.01025&3.14411&14.0&86.0\\
4.5&0.05852&0.0117&2.45327&2.87636&3.97500&8.4&58.0\\
4.5&0.00931&0.02793&4.29245&2.74634&3.47712&10.0&118.0\\
4.0&0.06473&0.00762&2.41937&2.84701&4.15708&3.64706&38.0\\
4.0&0.03891&0.01946&2.62902&3.08280&3.21988&14.0&86.0\\
\end{tabular}
\caption{Descriptors from the 41st to the 80th material. ({\it cont. 2})}
\label{c12}
\end{table}

\begin{table}
\begin{tabular}{cccccccccc}
material&$T_C$&$Z_R$&$r_R$&$r_R^{cov}$&$IP_R$&$\chi_R$&$S_{4f}$&$L_{4f}$&$J_{4f}$\\
Fe3Sm(PuNi3)&657&62&180&198&544.5&1.17&2.5&5&2.5\\
Mn23Sm6(Th6Mn23)&450&62&180&198&544.5&1.17&2.5&5&2.5\\
Ni2Sm(MgCu2)&22&62&180&198&544.5&1.17&2.5&5&2.5\\
Co2Tb(TbFe2)&230&65&177&194&565.8&1.2&3.0&3&6.0\\
Co5Tb(CaCu5)&979&65&177&194&565.8&1.2&3.0&3&6.0\\
CoTb3(Fe3C)&77&65&177&194&565.8&1.2&3.0&3&6.0\\
Fe17Tb2(Zn17Th2)&411&65&177&194&565.8&1.2&3.0&3&6.0\\
Fe2Tb(MgCu2)&701&65&177&194&565.8&1.2&3.0&3&6.0\\
Fe3Tb(PuNi3)&651&65&177&194&565.8&1.2&3.0&3&6.0\\
Mn23Tb6(Th6Mn23)&454&65&177&194&565.8&1.2&3.0&3&6.0\\
Mn2Tb(MgCu2)&48&65&177&194&565.8&1.2&3.0&3&6.0\\
Ni2Tb(MgCu2)&40&65&177&194&565.8&1.2&3.0&3&6.0\\
Ni5Tb(CaCu5)&23&65&177&194&565.8&1.2&3.0&3&6.0\\
Co2Tm(MgCu2)&4&69&176&190&596.7&1.25&1.0&5&6.0\\
Co3Tm(PuNi3)&370&69&176&190&596.7&1.25&1.0&5&6.0\\
Co7Tm2(Gd2Co7)&640&69&176&190&596.7&1.25&1.0&5&6.0\\
Fe17Tm2(Th2Ni17)&271&69&176&190&596.7&1.25&1.0&5&6.0\\
Fe2Tm(MgCu2)&562&69&176&190&596.7&1.25&1.0&5&6.0\\
NiTm(FeB-b)&7&69&176&190&596.7&1.25&1.0&5&6.0\\
Mn23Yb6(Th6Mn23)&406&70&176&187&603.4&1.1&0.5&3&3.5\\
\end{tabular}
\caption{Descriptors from the 81st to the 101st material.}
\label{c20}
\end{table}

\begin{table}
\begin{tabular}{cccccccccc}
$g_J$&$J_{4f} g_J$&$J_{4f} (g_J-1)$&$Z_T$&$r_T$&$r_T^{cov}$&$IP_T$&$\chi_T$&$S_3d$&$L_3d$\\
0.2857&0.71425&-1.78575&26.0&126.0&152.0&762.5&1.83&2.0&2.0\\
0.2857&0.71425&-1.78575&25.0&127.0&161.0&717.3&1.55&2.5&0.0\\
0.2857&0.71425&-1.78575&28.0&124.0&124.0&737.1&1.91&1.0&3.0\\
1.5&9.0&3.0&27.0&125.0&150.0&760.4&1.88&1.5&3.0\\
1.5&9.0&3.0&27.0&125.0&150.0&760.4&1.88&1.5&3.0\\
1.5&9.0&3.0&27.0&125.0&150.0&760.4&1.88&1.5&3.0\\
1.5&9.0&3.0&26.0&126.0&152.0&762.5&1.83&2.0&2.0\\
1.5&9.0&3.0&26.0&126.0&152.0&762.5&1.83&2.0&2.0\\
1.5&9.0&3.0&26.0&126.0&152.0&762.5&1.83&2.0&2.0\\
1.5&9.0&3.0&25.0&127.0&161.0&717.3&1.55&2.5&0.0\\
1.5&9.0&3.0&25.0&127.0&161.0&717.3&1.55&2.5&0.0\\
1.5&9.0&3.0&28.0&124.0&124.0&737.1&1.91&1.0&3.0\\
1.5&9.0&3.0&28.0&124.0&124.0&737.1&1.91&1.0&3.0\\
1.1667&7.0002&1.0002&27.0&125.0&150.0&760.4&1.88&1.5&3.0\\
1.1667&7.0002&1.0002&27.0&125.0&150.0&760.4&1.88&1.5&3.0\\
1.1667&7.0002&1.0002&27.0&125.0&150.0&760.4&1.88&1.5&3.0\\
1.1667&7.0002&1.0002&26.0&126.0&152.0&762.5&1.83&2.0&2.0\\
1.1667&7.0002&1.0002&26.0&126.0&152.0&762.5&1.83&2.0&2.0\\
1.1667&7.0002&1.0002&28.0&124.0&124.0&737.1&1.91&1.0&3.0\\
1.1429&4.00015&0.50015&25.0&127.0&161.0&717.3&1.55&2.5&0.0\\
\end{tabular}
\caption{Descriptors from the 81st to the 101st material. (cont.)}
\label{c21}
\end{table}

\begin{table}
\begin{tabular}{cccccccc}
$J_3d$&$C_T$&$C_R$&$d_{T-T}$&$d_{R-T}$&$d_{R-R}$&$N_{T-R}$&$N_{R-R}$\\
4.0&0.04687&0.01562&2.53719&2.96834&3.24242&10.22222&66.0\\
2.5&0.04632&0.01208&2.55984&3.08908&3.60866&5.47826&50.0\\
4.0&0.04234&0.02117&2.55619&2.99740&3.13068&14.0&86.0\\
4.5&0.04281&0.02141&2.54665&2.98621&3.11899&14.0&86.0\\
4.5&0.05916&0.01183&2.45023&2.85904&3.98000&8.4&58.0\\
4.5&0.00968&0.02903&4.25208&2.72082&3.41084&12.0&126.0\\
4.0&0.06599&0.00776&2.41336&2.82335&4.14675&3.64706&38.0\\
4.0&0.04038&0.02019&2.59685&3.04508&3.18048&14.0&86.0\\
4.0&0.04780&0.01593&2.52128&2.94881&3.22088&10.22222&66.0\\
2.5&0.04729&0.01234&2.54233&3.06794&3.58397&5.47826&50.0\\
2.5&0.03584&0.01792&2.70221&3.16862&3.30952&14.0&86.0\\
4.0&0.04332&0.02166&2.53675&2.97460&3.10687&14.0&86.0\\
4.0&0.06078&0.01216&2.43479&2.82555&3.96600&8.4&58.0\\
4.5&0.04436&0.02218&2.51663&2.95101&3.08223&14.0&86.0\\
4.5&0.05218&0.01739&2.46351&2.85910&3.11763&10.88889&78.0\\
4.5&0.05459&0.0156&2.43763&2.83463&3.13565&10.0&68.5\\
4.0&0.06686&0.00787&2.31784&2.80705&4.13900&4.35294&36.0\\
4.0&0.04199&0.02099&2.56326&3.00569&3.13934&14.0&86.0\\
4.0&0.02598&0.02598&2.42112&2.78329&3.50683&15.0&108.0\\
2.5&0.05080&0.01325&2.48225&2.99545&3.49928&8.6087&54.0\\
\end{tabular}
\caption{Descriptors from the 81st to the 101st material. (cont. 2)}
\label{c22}
\end{table}

\begin{table}
\begin{tabular}{cccccccccc}
material&$T_C$&$Z_R$&$r_R$&$r_R^{cov}$&$IP_R$&$\chi_R$&$S_{4f}$&$L_{4f}$&$J_{4f}$\\
Co17Ce2(Zn17Th2)&1090&58&181&204&534.4&1.12&0.5&3&2.5\\
Co5Ce(CaCu5)&662&58&181&204&534.4&1.12&0.5&3&2.5\\
Co7Ce2(Gd2Co7)&50&58&181&204&534.4&1.12&0.5&3&2.5\\
Fe17Ce2(Zn17Th2)&233&58&181&204&534.4&1.12&0.5&3&2.5\\
Fe2Ce(MgCu2)&235&58&181&204&534.4&1.12&0.5&3&2.5\\
Co2Dy(MgCu2)&141&66&178&192&573.0&1.22&2.5&5&7.5\\
Co3Dy(PuNi3)&450&66&178&192&573.0&1.22&2.5&5&7.5\\
Co5Dy(CaCu5)&977&66&178&192&573.0&1.22&2.5&5&7.5\\
Mn23Dy6(Th6Mn23)&443&66&178&192&573.0&1.22&2.5&5&7.5\\
Mn2Dy(MgZn2)&37&66&178&192&573.0&1.22&2.5&5&7.5\\
Ni17Dy2(Th2Ni17)&170&66&178&192&573.0&1.22&2.5&5&7.5\\
Ni2Dy(MgCu2)&28&66&178&192&573.0&1.22&2.5&5&7.5\\
Ni5Dy(CaCu5)&13&66&178&192&573.0&1.22&2.5&5&7.5\\
NiDy(FeB-b)&64&66&178&192&573.0&1.22&2.5&5&7.5\\
Co2Er(TbFe2)&39&68&176&189&589.3&1.24&1.5&6&7.5\\
Co3Er(PuNi3)&401&68&176&189&589.3&1.24&1.5&6&7.5\\
Co5Er(CaCu5)&1066&68&176&189&589.3&1.24&1.5&6&7.5\\
CoEr3(Fe3C)&7&68&176&189&589.3&1.24&1.5&6&7.5\\
Fe17Er2(Th2Ni17)&299&68&176&189&589.3&1.24&1.5&6&7.5\\
Fe23Er6(Th6Mn23)&498&68&176&189&589.3&1.24&1.5&6&7.5\\
Fe2Er(MgCu2)&587&68&176&189&589.3&1.24&1.5&6&7.5\\
Ni3Er(PuNi3)&66&68&176&189&589.3&1.24&1.5&6&7.5\\
Ni5Er(CaCu5)&11&68&176&189&589.3&1.24&1.5&6&7.5\\
Ni7Er2(Gd2Co7)&74&68&176&189&589.3&1.24&1.5&6&7.5\\
NiEr(FeB-b)&12&68&176&189&589.3&1.24&1.5&6&7.5\\
Ni2Eu(MgCu2)&139&63&180&198&547.1&1.2&3.0&3&0.0\\
Co17Gd2(Zn17Th2)&1209&64&180&196&593.4&1.2&3.5&0&3.5\\
Co2Gd(MgCu2)&405&64&180&196&593.4&1.2&3.5&0&3.5\\
Co3Gd(PuNi3)&631&64&180&196&593.4&1.2&3.5&0&3.5\\
Co3Gd4(Ho6Co4.5)&233&64&180&196&593.4&1.2&3.5&0&3.5\\
Co5Gd(CaCu5)&1002&64&180&196&593.4&1.2&3.5&0&3.5\\
CoGd3(Fe3C)&143&64&180&196&593.4&1.2&3.5&0&3.5\\
Fe17Gd2(Zn17Th2)&479&64&180&196&593.4&1.2&3.5&0&3.5\\
Fe23Gd6(Th6Mn23)&659&64&180&196&593.4&1.2&3.5&0&3.5\\
Fe2Gd(MgCu2)&814&64&180&196&593.4&1.2&3.5&0&3.5\\
Fe3Gd(PuNi3)&725&64&180&196&593.4&1.2&3.5&0&3.5\\
Fe5Gd(CaCu5)&470&64&180&196&593.4&1.2&3.5&0&3.5\\
Mn23Gd6(Th6Mn23)&465&64&180&196&593.4&1.2&3.5&0&3.5\\
Mn2Gd(MgCu2)&135&64&180&196&593.4&1.2&3.5&0&3.5\\
Ni17Gd2(Th2Ni17)&189&64&180&196&593.4&1.2&3.5&0&3.5\\
\end{tabular}
\caption{Descriptors from the 1st to the 40th material.}
\label{content00}
\end{table}

\begin{table}
\begin{tabular}{cccccccccc}
$g_J$&$J_{4f} g_J$&$J_{4f} (g_J-1)$&$Z_T$&$r_T$&$r_T^{cov}$&$IP_T$&$\chi_T$&$S_3d$&$L_3d$\\
0.8571&2.14275&-0.35725&27.0&125.0&150.0&760.4&1.88&1.5&3.0\\
0.8571&2.14275&-0.35725&27.0&125.0&150.0&760.4&1.88&1.5&3.0\\
0.8571&2.14275&-0.35725&27.0&125.0&150.0&760.4&1.88&1.5&3.0\\
0.8571&2.14275&-0.35725&26.0&126.0&152.0&762.5&1.83&2.0&2.0\\
0.8571&2.14275&-0.35725&26.0&126.0&152.0&762.5&1.83&2.0&2.0\\
1.3333&9.99975&2.49975&27.0&125.0&150.0&760.4&1.88&1.5&3.0\\
1.3333&9.99975&2.49975&27.0&125.0&150.0&760.4&1.88&1.5&3.0\\
1.3333&9.99975&2.49975&27.0&125.0&150.0&760.4&1.88&1.5&3.0\\
1.3333&9.99975&2.49975&25.0&127.0&161.0&717.3&1.55&2.5&0.0\\
1.3333&9.99975&2.49975&25.0&127.0&161.0&717.3&1.55&2.5&0.0\\
1.3333&9.99975&2.49975&28.0&124.0&124.0&737.1&1.91&1.0&3.0\\
1.3333&9.99975&2.49975&28.0&124.0&124.0&737.1&1.91&1.0&3.0\\
1.3333&9.99975&2.49975&28.0&124.0&124.0&737.1&1.91&1.0&3.0\\
1.3333&9.99975&2.49975&28.0&124.0&124.0&737.1&1.91&1.0&3.0\\
1.2&9.0&1.5&27.0&125.0&150.0&760.4&1.88&1.5&3.0\\
1.2&9.0&1.5&27.0&125.0&150.0&760.4&1.88&1.5&3.0\\
1.2&9.0&1.5&27.0&125.0&150.0&760.4&1.88&1.5&3.0\\
1.2&9.0&1.5&27.0&125.0&150.0&760.4&1.88&1.5&3.0\\
1.2&9.0&1.5&26.0&126.0&152.0&762.5&1.83&2.0&2.0\\
1.2&9.0&1.5&26.0&126.0&152.0&762.5&1.83&2.0&2.0\\
1.2&9.0&1.5&26.0&126.0&152.0&762.5&1.83&2.0&2.0\\
1.2&9.0&1.5&28.0&124.0&124.0&737.1&1.91&1.0&3.0\\
1.2&9.0&1.5&28.0&124.0&124.0&737.1&1.91&1.0&3.0\\
1.2&9.0&1.5&28.0&124.0&124.0&737.1&1.91&1.0&3.0\\
1.2&9.0&1.5&28.0&124.0&124.0&737.1&1.91&1.0&3.0\\
0.0&0.0&0.0&28.0&124.0&124.0&737.1&1.91&1.0&3.0\\
2.0&7.0&3.5&27.0&125.0&150.0&760.4&1.88&1.5&3.0\\
2.0&7.0&3.5&27.0&125.0&150.0&760.4&1.88&1.5&3.0\\
2.0&7.0&3.5&27.0&125.0&150.0&760.4&1.88&1.5&3.0\\
2.0&7.0&3.5&27.0&125.0&150.0&760.4&1.88&1.5&3.0\\
2.0&7.0&3.5&27.0&125.0&150.0&760.4&1.88&1.5&3.0\\
2.0&7.0&3.5&27.0&125.0&150.0&760.4&1.88&1.5&3.0\\
2.0&7.0&3.5&26.0&126.0&152.0&762.5&1.83&2.0&2.0\\
2.0&7.0&3.5&26.0&126.0&152.0&762.5&1.83&2.0&2.0\\
2.0&7.0&3.5&26.0&126.0&152.0&762.5&1.83&2.0&2.0\\
2.0&7.0&3.5&26.0&126.0&152.0&762.5&1.83&2.0&2.0\\
2.0&7.0&3.5&26.0&126.0&152.0&762.5&1.83&2.0&2.0\\
2.0&7.0&3.5&25.0&127.0&161.0&717.3&1.55&2.5&0.0\\
2.0&7.0&3.5&25.0&127.0&161.0&717.3&1.55&2.5&0.0\\
2.0&7.0&3.5&28.0&124.0&124.0&737.1&1.91&1.0&3.0\\
\end{tabular}
\caption{Descriptors from the 1st to the 40th material. (cont.)}
\label{content01}
\end{table}

\begin{table}
\begin{tabular}{cccccccc}
$J_3d$&$C_T$&$C_R$&$d_{T-T}$&$d_{R-T}$&$d_{R-R}$&$N_{T-R}$&$N_{R-R}$\\
4.5&0.06846&0.00805&2.37126&2.79635&4.07441&5.76471&38.0\\
4.5&0.05917&0.01183&2.46168&2.84518&4.01800&8.4&58.0\\
4.5&0.05449&0.01557&2.45226&2.83193&3.12910&10.0&68.5\\
4.0&0.06572&0.00773&2.40832&2.83201&4.13808&3.64706&38.0\\
4.0&0.04110&0.02055&2.58165&3.02725&3.16186&14.0&86.0\\
4.5&0.04294&0.02147&2.54417&2.98330&3.11596&14.0&86.0\\
4.5&0.05123&0.01708&2.47726&2.87697&3.13757&10.88889&78.0\\
4.5&0.08345&0.01192&1.54703&1.22008&3.98720&6.57143&58.0\\
2.5&0.04780&0.01247&2.53316&3.05688&3.57105&5.47826&50.0\\
2.5&0.03659&0.0183&2.68347&3.14665&3.28657&14.0&86.0\\
4.0&0.07398&0.0087&2.25952&2.60907&4.33000&4.35294&36.0\\
4.0&0.04366&0.02183&2.53003&2.96672&3.09864&14.0&86.0\\
4.0&0.06136&0.01227&2.43184&2.81112&3.96900&8.4&58.0\\
4.0&0.02502&0.02502&2.45561&2.81558&3.55253&11.0&106.0\\
4.5&0.04364&0.02182&2.53038&2.96714&3.09907&14.0&86.0\\
4.5&0.05188&0.01729&2.46788&2.86461&3.12373&10.88889&78.0\\
4.5&0.06033&0.01207&2.44450&2.82267&4.00400&8.4&58.0\\
4.5&0.01019&0.03056&4.19439&2.68390&3.32898&16.0&132.0\\
4.0&0.06635&0.00781&2.32148&2.81572&4.14550&3.64706&36.0\\
4.0&0.05332&0.01391&2.44254&2.94752&3.44330&8.6087&62.0\\
4.0&0.04162&0.02081&2.57069&3.01440&3.14844&14.0&86.0\\
4.0&0.05241&0.01747&2.46367&2.85376&3.11046&10.88889&78.0\\
4.0&0.06173&0.01235&2.42800&2.80361&3.96600&8.4&58.0\\
4.0&0.05579&0.01594&2.42823&2.81128&3.10758&10.0&68.5\\
4.0&0.02567&0.02567&2.43010&2.79557&3.52136&15.0&106.0\\
4.0&0.04056&0.02028&2.59296&3.04052&3.17572&14.0&86.0\\
4.5&0.06896&0.00811&2.36486&2.79001&4.06341&5.76471&38.0\\
4.5&0.04149&0.02074&2.57352&3.01771&3.15190&14.0&86.0\\
4.5&0.05047&0.01682&2.48879&2.89171&3.15397&10.22222&66.0\\
4.5&0.02116&0.02539&2.02400&2.85960&3.40595&12.6&107.0\\
4.5&0.05900&0.0118&2.44924&2.86539&3.97300&8.4&58.0\\
4.5&0.00941&0.02822&4.28765&2.74391&3.45177&10.0&122.0\\
4.0&0.06518&0.00767&2.42005&2.96247&4.17250&4.35294&36.0\\
4.0&0.05210&0.01359&2.46148&2.97038&3.47000&8.6087&62.0\\
4.0&0.03942&0.01971&2.61771&3.06954&3.20603&14.0&86.0\\
4.0&0.04746&0.01582&2.52733&2.95576&3.22844&10.22222&66.0\\
4.0&0.05992&0.01198&2.41500&2.78860&4.13000&8.4&58.0\\
2.5&0.04705&0.01227&2.54660&3.07310&3.59000&5.47826&50.0\\
2.5&0.03451&0.01725&2.73650&3.20883&3.35152&6.0&70.0\\
4.0&0.06915&0.00814&2.36257&2.72806&4.23700&4.35294&36.0\\
\end{tabular}
\caption{Descriptors from the 1st to the 40th material. ({\it cont. 2})}
\label{content02}
\end{table}

\begin{table}
\begin{tabular}{cccccccccc}
material&$T_C$&$Z_R$&$r_R$&$r_R^{cov}$&$IP_R$&$\chi_R$&$S_{4f}$&$L_{4f}$&$J_{4f}$\\
Ni2Gd(MgCu2)&77&64&180&196&593.4&1.2&3.5&0&3.5\\
Ni3Gd(PuNi3)&113&64&180&196&593.4&1.2&3.5&0&3.5\\
Ni5Gd(CaCu5)&32&64&180&196&593.4&1.2&3.5&0&3.5\\
NiGd(TlI)&71&64&180&196&593.4&1.2&3.5&0&3.5\\
Co2Ho(MgCu2)&83&67&176&192&581.0&1.23&2.0&6&8.0\\
Co5Ho(CaCu5)&1026&67&176&192&581.0&1.23&2.0&6&8.0\\
CoHo3(Fe3C)&10&67&176&192&581.0&1.23&2.0&6&8.0\\
Fe17Ho2(Th2Ni17)&335&67&176&192&581.0&1.23&2.0&6&8.0\\
Fe23Ho6(Th6Mn23)&507&67&176&192&581.0&1.23&2.0&6&8.0\\
Fe2Ho(MgCu2)&560&67&176&192&581.0&1.23&2.0&6&8.0\\
Mn2Ho(MgCu2)&25&67&176&192&581.0&1.23&2.0&6&8.0\\
Ni2Ho(MgCu2)&16&67&176&192&581.0&1.23&2.0&6&8.0\\
Ni5Ho(CaCu5)&14&67&176&192&581.0&1.23&2.0&6&8.0\\
NiHo(FeB-b)&38&67&176&192&581.0&1.23&2.0&6&8.0\\
Co13La(NaZn13)&1298&57&187&207&538.1&1.1&0.0&0&0.0\\
Co5La(CaCu5)&835&57&187&207&538.1&1.1&0.0&0&0.0\\
Fe2Lu(MgCu2)&580&71&174&187&523.5&1.27&0.0&0&0.0\\
Co2Nd(MgCu2)&108&60&181&201&533.1&1.14&1.5&6&4.5\\
Co3Nd(PuNi3)&381&60&181&201&533.1&1.14&1.5&6&4.5\\
Co5Nd(CaCu5)&910&60&181&201&533.1&1.14&1.5&6&4.5\\
CoNd3(Fe3C)&27&60&181&201&533.1&1.14&1.5&6&4.5\\
Fe17Nd2(Zn17Th2)&327&60&181&201&533.1&1.14&1.5&6&4.5\\
Fe17Nd5(Nd5Fe17)&303&60&181&201&533.1&1.14&1.5&6&4.5\\
Fe2Nd(MgCu2)&453&60&181&201&533.1&1.14&1.5&6&4.5\\
Mn23Nd6(Th6Mn23)&438&60&181&201&533.1&1.14&1.5&6&4.5\\
Ni2Nd(MgCu2)&9&60&181&201&533.1&1.14&1.5&6&4.5\\
Ni5Nd(CaCu5)&7&60&181&201&533.1&1.14&1.5&6&4.5\\
Co2Pr(MgCu2)&45&59&182&203&527.0&1.13&1.0&5&4.0\\
Co5Pr(CaCu5)&931&59&182&203&527.0&1.13&1.0&5&4.0\\
CoPr3(Fe3C)&14&59&182&203&527.0&1.13&1.0&5&4.0\\
Fe17Pr2(Zn17Th2)&280&59&182&203&527.0&1.13&1.0&5&4.0\\
Mn23Pr6(Th6Mn23)&448&59&182&203&527.0&1.13&1.0&5&4.0\\
Ni5Pr(CaCu5)&0&59&182&203&527.0&1.13&1.0&5&4.0\\
Co17Sm2(Zn17Th2)&1193&62&180&198&544.5&1.17&2.5&5&2.5\\
Co2Sm(MgCu2)&230&62&180&198&544.5&1.17&2.5&5&2.5\\
Co5Sm(CaCu5)&1016&62&180&198&544.5&1.17&2.5&5&2.5\\
CoSm3(Fe3C)&78&62&180&198&544.5&1.17&2.5&5&2.5\\
Fe17Sm2(Zn17Th2)&395&62&180&198&544.5&1.17&2.5&5&2.5\\
Fe2Sm(MgCu2)&676&62&180&198&544.5&1.17&2.5&5&2.5\\
Fe3Sm(PuNi3)&657&62&180&198&544.5&1.17&2.5&5&2.5\\
\end{tabular}
\caption{Descriptors from the 41st to the 80th material. }
\label{content10}

\end{table}

\begin{table}
\begin{tabular}{cccccccccc}
$g_J$&$J_{4f} g_J$&$J_{4f} (g_J-1)$&$Z_T$&$r_T$&$r_T^{cov}$&$IP_T$&$\chi_T$&$S_3d$&$L_3d$\\
2.0&7.0&3.5&28.0&124.0&124.0&737.1&1.91&1.0&3.0\\
2.0&7.0&3.5&28.0&124.0&124.0&737.1&1.91&1.0&3.0\\
2.0&7.0&3.5&28.0&124.0&124.0&737.1&1.91&1.0&3.0\\
2.0&7.0&3.5&28.0&124.0&124.0&737.1&1.91&1.0&3.0\\
1.25&10.0&2.0&27.0&125.0&150.0&760.4&1.88&1.5&3.0\\
1.25&10.0&2.0&27.0&125.0&150.0&760.4&1.88&1.5&3.0\\
1.25&10.0&2.0&27.0&125.0&150.0&760.4&1.88&1.5&3.0\\
1.25&10.0&2.0&26.0&126.0&152.0&762.5&1.83&2.0&2.0\\
1.25&10.0&2.0&26.0&126.0&152.0&762.5&1.83&2.0&2.0\\
1.25&10.0&2.0&26.0&126.0&152.0&762.5&1.83&2.0&2.0\\
1.25&10.0&2.0&25.0&127.0&161.0&717.3&1.55&2.5&0.0\\
1.25&10.0&2.0&28.0&124.0&124.0&737.1&1.91&1.0&3.0\\
1.25&10.0&2.0&28.0&124.0&124.0&737.1&1.91&1.0&3.0\\
1.25&10.0&2.0&28.0&124.0&124.0&737.1&1.91&1.0&3.0\\
0.0&0.0&0.0&27.0&125.0&150.0&760.4&1.88&1.5&3.0\\
0.0&0.0&0.0&27.0&125.0&150.0&760.4&1.88&1.5&3.0\\
0.0&0.0&0.0&26.0&126.0&152.0&762.5&1.83&2.0&2.0\\
0.7273&3.27285&-1.22715&27.0&125.0&150.0&760.4&1.88&1.5&3.0\\
0.7273&3.27285&-1.22715&27.0&125.0&150.0&760.4&1.88&1.5&3.0\\
0.7273&3.27285&-1.22715&27.0&125.0&150.0&760.4&1.88&1.5&3.0\\
0.7273&3.27285&-1.22715&27.0&125.0&150.0&760.4&1.88&1.5&3.0\\
0.7273&3.27285&-1.22715&26.0&126.0&152.0&762.5&1.83&2.0&2.0\\
0.7273&3.27285&-1.22715&26.0&126.0&152.0&762.5&1.83&2.0&2.0\\
0.7273&3.27285&-1.22715&26.0&126.0&152.0&762.5&1.83&2.0&2.0\\
0.7273&3.27285&-1.22715&25.0&127.0&161.0&717.3&1.55&2.5&0.0\\
0.7273&3.27285&-1.22715&28.0&124.0&124.0&737.1&1.91&1.0&3.0\\
0.7273&3.27285&-1.22715&28.0&124.0&124.0&737.1&1.91&1.0&3.0\\
0.8&3.2&-0.8&27.0&125.0&150.0&760.4&1.88&1.5&3.0\\
0.8&3.2&-0.8&27.0&125.0&150.0&760.4&1.88&1.5&3.0\\
0.8&3.2&-0.8&27.0&125.0&150.0&760.4&1.88&1.5&3.0\\
0.8&3.2&-0.8&26.0&126.0&152.0&762.5&1.83&2.0&2.0\\
0.8&3.2&-0.8&25.0&127.0&161.0&717.3&1.55&2.5&0.0\\
0.8&3.2&-0.8&28.0&124.0&124.0&737.1&1.91&1.0&3.0\\
0.2857&0.71425&-1.78575&27.0&125.0&150.0&760.4&1.88&1.5&3.0\\
0.2857&0.71425&-1.78575&27.0&125.0&150.0&760.4&1.88&1.5&3.0\\
0.2857&0.71425&-1.78575&27.0&125.0&150.0&760.4&1.88&1.5&3.0\\
0.2857&0.71425&-1.78575&27.0&125.0&150.0&760.4&1.88&1.5&3.0\\
0.2857&0.71425&-1.78575&26.0&126.0&152.0&762.5&1.83&2.0&2.0\\
0.2857&0.71425&-1.78575&26.0&126.0&152.0&762.5&1.83&2.0&2.0\\
0.2857&0.71425&-1.78575&26.0&126.0&152.0&762.5&1.83&2.0&2.0\\
\end{tabular}
\caption{Descriptors from the 41st to the 80th material. (cont.) }
\label{content11}
\end{table}

\begin{table}
\begin{tabular}{cccccccc}
$J_3d$&$C_T$&$C_R$&$d_{T-T}$&$d_{R-T}$&$d_{R-R}$&$N_{T-R}$&$N_{R-R}$\\
4.0&0.04265&0.02133&2.54983&2.98994&3.12289&14.0&86.0\\
4.0&0.05057&0.01686&2.49359&2.88764&3.14721&10.22222&66.0\\
4.0&0.06042&0.01208&2.43690&2.83421&3.96500&8.4&58.0\\
4.0&0.02420&0.0242&2.58495&2.89772&3.58845&11.0&92.0\\
4.5&0.04333&0.02167&2.53639&2.97418&3.10643&14.0&86.0\\
4.5&0.05994&0.01199&2.44444&2.84056&3.97900&8.4&58.0\\
4.5&0.01001&0.03003&4.20705&2.69174&3.36592&14.0&131.33333\\
4.0&0.06652&0.00783&2.32484&2.81005&4.15150&3.64706&36.0\\
4.0&0.05259&0.01372&2.45374&2.96104&3.45909&8.6087&62.0\\
4.0&0.04058&0.02029&2.59261&3.04010&3.17528&14.0&86.0\\
2.5&0.03852&0.01926&2.63800&3.09333&3.23088&14.0&86.0\\
4.0&0.04410&0.02205&2.52154&2.95677&3.08825&14.0&86.0\\
4.0&0.06135&0.01227&2.43006&2.81343&3.96300&8.4&58.0\\
4.0&0.02533&0.02533&2.44355&2.80573&3.53764&11.0&106.0\\
4.5&0.07100&0.00546&2.37741&3.29960&5.67850&2.46154&26.0\\
4.5&0.05567&0.01113&2.47327&2.95084&3.97000&8.4&52.0\\
4.0&0.04234&0.02117&2.55619&2.99740&3.13068&14.0&86.0\\
4.5&0.04096&0.02048&2.58448&3.03057&3.16532&14.0&86.0\\
4.5&0.04914&0.01638&2.51634&2.91590&3.17848&10.22222&66.0\\
4.5&0.05728&0.01146&2.46505&2.90407&3.98400&8.4&52.0\\
4.5&0.00897&0.02692&4.33408&2.77265&3.53652&10.0&115.33333\\
4.0&0.06421&0.00755&2.39377&3.07380&3.91324&4.35294&38.0\\
4.0&0.04675&0.01375&2.33911&2.96841&3.25413&7.52941&56.2\\
4.0&0.03854&0.01927&2.63751&3.09275&3.23027&14.0&86.0\\
2.5&0.04510&0.01177&2.58265&3.11660&3.64082&5.47826&50.0\\
4.0&0.04167&0.02083&2.56973&3.01328&3.14727&14.0&86.0\\
4.0&0.05919&0.01184&2.44789&2.86077&3.97300&8.4&58.0\\
4.5&0.04093&0.02046&2.58518&3.03140&3.16619&14.0&86.0\\
4.5&0.05774&0.01155&2.46101&2.89310&3.98200&8.4&52.0\\
4.5&0.00890&0.0267&4.33876&2.77540&3.55680&10.0&115.33333\\
4.0&0.06417&0.00755&2.41782&2.86035&4.15442&3.64706&38.0\\
2.5&0.04465&0.01165&2.59140&3.12717&3.65316&5.47826&50.0\\
4.0&0.05914&0.01183&2.44823&2.86193&3.97300&8.4&58.0\\
4.5&0.06835&0.00804&2.36874&2.80001&4.07008&5.76471&38.0\\
4.5&0.04180&0.0209&2.56715&3.01025&3.14411&14.0&86.0\\
4.5&0.05852&0.0117&2.45327&2.87636&3.97500&8.4&58.0\\
4.5&0.00931&0.02793&4.29245&2.74634&3.47712&10.0&118.0\\
4.0&0.06473&0.00762&2.41937&2.84701&4.15708&3.64706&38.0\\
4.0&0.03891&0.01946&2.62902&3.08280&3.21988&14.0&86.0\\
4.0&0.04687&0.01562&2.53719&2.96834&3.24242&10.22222&66.0\\
\end{tabular}
\caption{Descriptors from the 41st to the 80th material. ({\it cont. 2}) }
\label{content12}
\end{table}

\begin{table}
\begin{tabular}{cccccccccc}
material&$T_C$&$Z_R$&$r_R$&$r_R^{cov}$&$IP_R$&$\chi_R$&$S_{4f}$&$L_{4f}$&$J_{4f}$\\
Mn23Sm6(Th6Mn23)&450&62&180&198&544.5&1.17&2.5&5&2.5\\
Ni2Sm(MgCu2)&22&62&180&198&544.5&1.17&2.5&5&2.5\\
Co2Tb(TbFe2)&230&65&177&194&565.8&1.2&3.0&3&6.0\\
Co5Tb(CaCu5)&979&65&177&194&565.8&1.2&3.0&3&6.0\\
CoTb3(Fe3C)&77&65&177&194&565.8&1.2&3.0&3&6.0\\
Fe17Tb2(Zn17Th2)&411&65&177&194&565.8&1.2&3.0&3&6.0\\
Fe2Tb(MgCu2)&701&65&177&194&565.8&1.2&3.0&3&6.0\\
Fe3Tb(PuNi3)&651&65&177&194&565.8&1.2&3.0&3&6.0\\
Mn23Tb6(Th6Mn23)&454&65&177&194&565.8&1.2&3.0&3&6.0\\
Mn2Tb(MgCu2)&48&65&177&194&565.8&1.2&3.0&3&6.0\\
Ni2Tb(MgCu2)&40&65&177&194&565.8&1.2&3.0&3&6.0\\
Ni5Tb(CaCu5)&23&65&177&194&565.8&1.2&3.0&3&6.0\\
Co2Tm(MgCu2)&4&69&176&190&596.7&1.25&1.0&5&6.0\\
Co3Tm(PuNi3)&370&69&176&190&596.7&1.25&1.0&5&6.0\\
Co7Tm2(Gd2Co7)&640&69&176&190&596.7&1.25&1.0&5&6.0\\
Fe17Tm2(Th2Ni17)&271&69&176&190&596.7&1.25&1.0&5&6.0\\
Fe2Tm(MgCu2)&562&69&176&190&596.7&1.25&1.0&5&6.0\\
NiTm(FeB-b)&7&69&176&190&596.7&1.25&1.0&5&6.0\\
Mn23Yb6(Th6Mn23)&406&70&176&187&603.4&1.1&0.5&3&3.5\\
\end{tabular}
\caption{Descriptors from the 81st to the 100th material.}
\label{content20}
\end{table}

\begin{table}
\begin{tabular}{cccccccccc}
$g_J$&$J_{4f} g_J$&$J_{4f} (g_J-1)$&$Z_T$&$r_T$&$r_T^{cov}$&$IP_T$&$\chi_T$&$S_3d$&$L_3d$\\
0.2857&0.71425&-1.78575&25.0&127.0&161.0&717.3&1.55&2.5&0.0\\
0.2857&0.71425&-1.78575&28.0&124.0&124.0&737.1&1.91&1.0&3.0\\
1.5&9.0&3.0&27.0&125.0&150.0&760.4&1.88&1.5&3.0\\
1.5&9.0&3.0&27.0&125.0&150.0&760.4&1.88&1.5&3.0\\
1.5&9.0&3.0&27.0&125.0&150.0&760.4&1.88&1.5&3.0\\
1.5&9.0&3.0&26.0&126.0&152.0&762.5&1.83&2.0&2.0\\
1.5&9.0&3.0&26.0&126.0&152.0&762.5&1.83&2.0&2.0\\
1.5&9.0&3.0&26.0&126.0&152.0&762.5&1.83&2.0&2.0\\
1.5&9.0&3.0&25.0&127.0&161.0&717.3&1.55&2.5&0.0\\
1.5&9.0&3.0&25.0&127.0&161.0&717.3&1.55&2.5&0.0\\
1.5&9.0&3.0&28.0&124.0&124.0&737.1&1.91&1.0&3.0\\
1.5&9.0&3.0&28.0&124.0&124.0&737.1&1.91&1.0&3.0\\
1.1667&7.0002&1.0002&27.0&125.0&150.0&760.4&1.88&1.5&3.0\\
1.1667&7.0002&1.0002&27.0&125.0&150.0&760.4&1.88&1.5&3.0\\
1.1667&7.0002&1.0002&27.0&125.0&150.0&760.4&1.88&1.5&3.0\\
1.1667&7.0002&1.0002&26.0&126.0&152.0&762.5&1.83&2.0&2.0\\
1.1667&7.0002&1.0002&26.0&126.0&152.0&762.5&1.83&2.0&2.0\\
1.1667&7.0002&1.0002&28.0&124.0&124.0&737.1&1.91&1.0&3.0\\
1.1429&4.00015&0.50015&25.0&127.0&161.0&717.3&1.55&2.5&0.0\\
\end{tabular}
\caption{Descriptors from the 81st to the 100th materials. (cont.) }
\label{content21}
\end{table}

\begin{table}
\begin{tabular}{cccccccc}
$J_3d$&$C_T$&$C_R$&$d_{T-T}$&$d_{R-T}$&$d_{R-R}$&$N_{T-R}$&$N_{R-R}$\\
2.5&0.04632&0.01208&2.55984&3.08908&3.60866&5.47826&50.0\\
4.0&0.04234&0.02117&2.55619&2.99740&3.13068&14.0&86.0\\
4.5&0.04281&0.02141&2.54665&2.98621&3.11899&14.0&86.0\\
4.5&0.05916&0.01183&2.45023&2.85904&3.98000&8.4&58.0\\
4.5&0.00968&0.02903&4.25208&2.72082&3.41084&12.0&126.0\\
4.0&0.06599&0.00776&2.41336&2.82335&4.14675&3.64706&38.0\\
4.0&0.04038&0.02019&2.59685&3.04508&3.18048&14.0&86.0\\
4.0&0.04780&0.01593&2.52128&2.94881&3.22088&10.22222&66.0\\
2.5&0.04729&0.01234&2.54233&3.06794&3.58397&5.47826&50.0\\
2.5&0.03584&0.01792&2.70221&3.16862&3.30952&14.0&86.0\\
4.0&0.04332&0.02166&2.53675&2.97460&3.10687&14.0&86.0\\
4.0&0.06078&0.01216&2.43479&2.82555&3.96600&8.4&58.0\\
4.5&0.04436&0.02218&2.51663&2.95101&3.08223&14.0&86.0\\
4.5&0.05218&0.01739&2.46351&2.85910&3.11763&10.88889&78.0\\
4.5&0.05459&0.0156&2.43763&2.83463&3.13565&10.0&68.5\\
4.0&0.06686&0.00787&2.31784&2.80705&4.13900&4.35294&36.0\\
4.0&0.04199&0.02099&2.56326&3.00569&3.13934&14.0&86.0\\
4.0&0.02598&0.02598&2.42112&2.78329&3.50683&15.0&108.0\\
2.5&0.05080&0.01325&2.48225&2.99545&3.49928&8.6087&54.0\\
\end{tabular}
\caption{Descriptors from the 81st to the 100th material.  ({\it cont. 2}) }
\label{content22}
\end{table}